\input harvmac

\input amssym
\input epsf


\newfam\frakfam
\font\teneufm=eufm10
\font\seveneufm=eufm7
\font\fiveeufm=eufm5
\textfont\frakfam=\teneufm
\scriptfont\frakfam=\seveneufm
\scriptscriptfont\frakfam=\fiveeufm


\def\bb{
\font\tenmsb=msbm10
\font\sevenmsb=msbm7
\font\fivemsb=msbm5
\textfont1=\tenmsb
\scriptfont1=\sevenmsb
\scriptscriptfont1=\fivemsb
}


\newfam\dsromfam
\font\tendsrom=dsrom10
\textfont\dsromfam=\tendsrom
\def\ds{\fam\dsromfam \tendsrom}


\newfam\mbffam
\font\tenmbf=cmmib10
\font\sevenmbf=cmmib7
\font\fivembf=cmmib5
\textfont\mbffam=\tenmbf
\scriptfont\mbffam=\sevenmbf
\scriptscriptfont\mbffam=\fivembf


\newfam\mbfcalfam
\font\tenmbfcal=cmbsy10
\font\sevenmbfcal=cmbsy7
\font\fivembfcal=cmbsy5
\textfont\mbfcalfam=\tenmbfcal
\scriptfont\mbfcalfam=\sevenmbfcal
\scriptscriptfont\mbfcalfam=\fivembfcal


\newfam\mscrfam
\font\tenmscr=rsfs10
\font\sevenmscr=rsfs7
\font\fivemscr=rsfs5
\textfont\mscrfam=\tenmscr
\scriptfont\mscrfam=\sevenmscr
\scriptscriptfont\mscrfam=\fivemscr




\def\tilde{\widetilde}

\def\bar{\overline}
\def\b{\bar}
\def\bsq#1{{{\b{#1}}^{\lower 2.5pt\hbox{$\scriptstyle 2$}}}}
\def\bexp#1#2{{{\b{#1}}^{\lower 2.5pt\hbox{$\scriptstyle #2$}}}}
\def\dotexp#1#2{{{#1}^{\lower 2.5pt\hbox{$\scriptstyle #2$}}}}

\def\IL{\relax{\rm I\kern-.18em L}}
\def\IH{\relax{\rm I\kern-.18em H}}
\def\IR{\relax{\rm I\kern-.18em R}}
\def\IC{\relax{\rm I\kern-0.54 em C}}

\def\rt2{\sqrt{2}}
\def\half {{1 \over 2}}

\def\mod{{\rm mod}}

\def\Tr{\mathop{\rm Tr}}


\font\tenbifull=cmmib10
\font\tenbimed=cmmib7
\font\tenbismall=cmmib5
\textfont9=\tenbifull \scriptfont9=\tenbimed
\scriptscriptfont9=\tenbismall

\mathchardef\bbGamma="7000
\mathchardef\bbDelta="7001
\mathchardef\bbPhi="7002
\mathchardef\bbAlpha="7003
\mathchardef\bbXi="7004
\mathchardef\bbPi="7005
\mathchardef\bbSigma="7006
\mathchardef\bbUpsilon="7007
\mathchardef\bbTheta="7008
\mathchardef\bbPsi="7009
\mathchardef\bbOmega="700A
\mathchardef\bbalpha="710B
\mathchardef\bbbeta="710C
\mathchardef\bbgamma="710D
\mathchardef\bbdelta="710E
\mathchardef\bbepsilon="710F
\mathchardef\bbzeta="7110
\mathchardef\bbeta="7111
\mathchardef\bbtheta="7112
\mathchardef\bbiota="7113
\mathchardef\bbkappa="7114
\mathchardef\bblambda="7115
\mathchardef\bbmu="7116
\mathchardef\bbnu="7117
\mathchardef\bbxi="7118
\mathchardef\bbpi="7119
\mathchardef\bbrho="711A
\mathchardef\bbsigma="711B
\mathchardef\bbtau="711C
\mathchardef\bbupsilon="711D
\mathchardef\bbphi="711E
\mathchardef\bbchi="711F
\mathchardef\bbpsi="7120
\mathchardef\bbomega="7121
\mathchardef\bbvarepsilon="7122
\mathchardef\bbvartheta="7123
\mathchardef\bbvarpi="7124
\mathchardef\bbvarrho="7125
\mathchardef\bbvarsigma="7126
\mathchardef\bbvarphi="7127

\def\IL{\relax{\rm I\kern-.18em L}}
\def\IH{\relax{\rm I\kern-.18em H}}
\def\IR{\relax{\rm I\kern-.18em R}}
\def\IC{\relax{\rm I\kern-0.54 em C}}





\def\CN{{\cal N}}


\def\1{{\ds 1}}
\def\R{\hbox{$\bb R$}}

\def\Z{\hbox{$\bb Z$}}


\noblackbox

\def\unit{\relax{\rm 1\kern-.26em I}}
\def\nada{\relax{\rm 0\kern-.30em l}}
\def\tilde{\widetilde}

\def\mod{{\rm mod}}

\noblackbox
\def\IL{\relax{\rm I\kern-.18em L}}
\def\IH{\relax{\rm I\kern-.18em H}}
\def\IR{\relax{\rm I\kern-.18em R}}
\def\IC{\relax\hbox{$\inbar\kern-.3em{\rm C}$}}
\def\IZ{\relax\ifmmode\mathchoice
{\hbox{\cmss Z\kern-.4em Z}}{\hbox{\cmss Z\kern-.4em Z}} {\lower.9pt\hbox{\cmsss Z\kern-.4em Z}}
{\lower1.2pt\hbox{\cmsss Z\kern-.4em Z}}\else{\cmss Z\kern-.4em Z}\fi}

\def\CN {{\cal N}}

\def\partialslash{\not{\hbox{\kern-2pt $\partial$}}}


\def\CN {{\cal N}}

\def\Tr{{\rm Tr}}

\font\manual=manfnt \def\dbend{\lower3.5pt\hbox{\manual\char127}}

\def\IZ{\relax\ifmmode\mathchoice
{\hbox{\cmss Z\kern-.4em Z}}{\hbox{\cmss Z\kern-.4em Z}} {\lower.9pt\hbox{\cmsss Z\kern-.4em Z}}
{\lower1.2pt\hbox{\cmsss Z\kern-.4em Z}}\else{\cmss Z\kern-.4em Z}\fi}
\def\half {{1\over 2}}

\def\bar{\overline}

\def\rt2{\sqrt{2}}
\def\irt2{{1\over\sqrt{2}}}

\def\slashchar#1{\setbox0=\hbox{$#1$}           
   \dimen0=\wd0                                 
   \setbox1=\hbox{/} \dimen1=\wd1               
   \ifdim\dimen0>\dimen1                        
      \rlap{\hbox to \dimen0{\hfil/\hfil}}      
      #1                                        
   \else                                        
      \rlap{\hbox to \dimen1{\hfil$#1$\hfil}}   
      /                                         
   \fi}

\def\foursqr#1#2{{\vcenter{\vbox{
    \hrule height.#2pt
    \hbox{\vrule width.#2pt height#1pt \kern#1pt
    \vrule width.#2pt}
    \hrule height.#2pt
    \hrule height.#2pt
    \hbox{\vrule width.#2pt height#1pt \kern#1pt
    \vrule width.#2pt}
    \hrule height.#2pt
        \hrule height.#2pt
    \hbox{\vrule width.#2pt height#1pt \kern#1pt
    \vrule width.#2pt}
    \hrule height.#2pt
        \hrule height.#2pt
    \hbox{\vrule width.#2pt height#1pt \kern#1pt
    \vrule width.#2pt}
    \hrule height.#2pt}}}}
\def\psqr#1#2{{\vcenter{\vbox{\hrule height.#2pt
    \hbox{\vrule width.#2pt height#1pt \kern#1pt
    \vrule width.#2pt}
    \hrule height.#2pt \hrule height.#2pt
    \hbox{\vrule width.#2pt height#1pt \kern#1pt
    \vrule width.#2pt}
    \hrule height.#2pt}}}}
\def\sqr#1#2{{\vcenter{\vbox{\hrule height.#2pt
    \hbox{\vrule width.#2pt height#1pt \kern#1pt
    \vrule width.#2pt}
    \hrule height.#2pt}}}}
\def\square{\mathchoice\sqr65\sqr65\sqr{2.1}3\sqr{1.5}3}

\def\figin{\epsfcheck\figin}\def\figins{\epsfcheck\figins}
\def\epsfcheck{\ifx\epsfbox\UnDeFiNeD
\message{(NO epsf.tex, FIGURES WILL BE IGNORED)}
\gdef\figin##1{\vskip2in}\gdef\figins##1{\hskip.5in}
\else\message{(FIGURES WILL BE INCLUDED)}%
\gdef\figin##1{##1}\gdef\figins##1{##1}\fi}
\def\DefWarn#1{}
\def\figinsert{\goodbreak\midinsert}
\def\ifig#1#2#3{\DefWarn#1\xdef#1{fig.~\the\figno}
\writedef{#1\leftbracket fig.\noexpand~\the\figno}%
\figinsert\figin{\centerline{#3}}\medskip\centerline{\vbox{\baselineskip12pt \advance\hsize by
-1truein\noindent\footnotefont{\bf Fig.~\the\figno:\ } \it#2}}
\bigskip\endinsert\global\advance\figno by1}


\lref\FestucciaWS{
  G.~Festuccia and N.~Seiberg,
  ``Rigid Supersymmetric Theories in Curved Superspace,''
JHEP {\bf 1106}, 114 (2011).
[arXiv:1105.0689 [hep-th]].
}

\lref\MartelliFU{
  D.~Martelli, A.~Passias and J.~Sparks,
  ``The Gravity dual of supersymmetric gauge theories on a squashed three-sphere,''
Nucl.\ Phys.\ B {\bf 864}, 840 (2012).
[arXiv:1110.6400 [hep-th]].
}

\lref\ClossetVG{
  C.~Closset, T.~T.~Dumitrescu, G.~Festuccia, Z.~Komargodski and N.~Seiberg,
  ``Contact Terms, Unitarity, and F-Maximization in Three-Dimensional Superconformal Theories,''
JHEP {\bf 1210}, 053 (2012).
[arXiv:1205.4142 [hep-th]].
}

\lref\ClossetVP{
  C.~Closset, T.~T.~Dumitrescu, G.~Festuccia, Z.~Komargodski and N.~Seiberg,
  ``Comments on Chern-Simons Contact Terms in Three Dimensions,''
JHEP {\bf 1209}, 091 (2012).
[arXiv:1206.5218 [hep-th]].
}

\lref\KomargodskiRB{
  Z.~Komargodski and N.~Seiberg,
  ``Comments on Supercurrent Multiplets, Supersymmetric Field Theories and Supergravity,''
JHEP {\bf 1007}, 017 (2010).
[arXiv:1002.2228 [hep-th]].
}

\lref\SohniusTP{
  M.~F.~Sohnius and P.~C.~West,
  ``An Alternative Minimal Off-Shell Version of N=1 Supergravity,''
Phys.\ Lett.\ B {\bf 105}, 353 (1981).
}

\lref\DolanRP{
  F.~A.~H.~Dolan, V.~P.~Spiridonov and G.~S.~Vartanov,
  ``From 4d superconformal indices to 3d partition functions,''
Phys.\ Lett.\ B {\bf 704}, 234 (2011).
[arXiv:1104.1787 [hep-th]].
}

\lref\DeserYX{
  S.~Deser and A.~Schwimmer,
  ``Geometric classification of conformal anomalies in arbitrary dimensions,''
Phys.\ Lett.\ B {\bf 309}, 279 (1993).
[hep-th/9302047].
}

\lref\GaddeIA{
  A.~Gadde and W.~Yan,
  ``Reducing the 4d Index to the $S^3$ Partition Function,''
JHEP {\bf 1212}, 003 (2012).
[arXiv:1104.2592 [hep-th]].
}

\lref\ImamuraUW{
  Y.~Imamura,
  ``Relation between the 4d superconformal index and the $S^3$ partition function,''
JHEP {\bf 1109}, 133 (2011).
[arXiv:1104.4482 [hep-th]].
}

\lref\DumitrescuHA{
  T.~T.~Dumitrescu, G.~Festuccia and N.~Seiberg,
  ``Exploring Curved Superspace,''
JHEP {\bf 1208}, 141 (2012).
[arXiv:1205.1115 [hep-th]].
}

\lref\DumitrescuAT{
  T.~T.~Dumitrescu and G.~Festuccia,
  ``Exploring Curved Superspace (II),''
JHEP {\bf 1301}, 072 (2013).
[arXiv:1209.5408 [hep-th]].
}

\lref\DumitrescuIU{
  T.~T.~Dumitrescu and N.~Seiberg,
  ``Supercurrents and Brane Currents in Diverse Dimensions,''
JHEP {\bf 1107}, 095 (2011).
[arXiv:1106.0031 [hep-th]].
}

\lref\SohniusFW{
  M.~Sohnius and P.~C.~West,
  ``The Tensor Calculus And Matter Coupling Of The Alternative Minimal Auxiliary Field Formulation Of N=1 Supergravity,''
Nucl.\ Phys.\ B {\bf 198}, 493 (1982).
}

\lref\KuzenkoXG{
  S.~M.~Kuzenko, U.~Lindstrom and G.~Tartaglino-Mazzucchelli,
  ``Off-shell supergravity-matter couplings in three dimensions,''
JHEP {\bf 1103}, 120 (2011).
[arXiv:1101.4013 [hep-th]].
}

\lref\KuzenkoBC{
  S.~M.~Kuzenko, U.~Lindstrom and G.~Tartaglino-Mazzucchelli,
  ``Three-dimensional (p,q) AdS superspaces and matter couplings,''
JHEP {\bf 1208}, 024 (2012).
[arXiv:1205.4622 [hep-th]].
}

\lref\WittenEV{
  E.~Witten,
  ``Supersymmetric Yang-Mills theory on a four manifold,''
J.\ Math.\ Phys.\  {\bf 35}, 5101 (1994).
[hep-th/9403195].
}

\lref\KlareGN{
  C.~Klare, A.~Tomasiello and A.~Zaffaroni,
  ``Supersymmetry on Curved Spaces and Holography,''
JHEP {\bf 1208}, 061 (2012).
[arXiv:1205.1062 [hep-th]].
}

\lref\KomargodskiPC{
  Z.~Komargodski and N.~Seiberg,
  ``Comments on the Fayet-Iliopoulos Term in Field Theory and Supergravity,''
JHEP {\bf 0906}, 007 (2009).
[arXiv:0904.1159 [hep-th]].
}

\lref\ImamuraSU{
  Y.~Imamura and S.~Yokoyama,
  ``Index for three dimensional superconformal field theories with general R-charge assignments,''
JHEP {\bf 1104}, 007 (2011).
[arXiv:1101.0557 [hep-th]].
}

\lref\HoweZM{
  P.~S.~Howe, J.~M.~Izquierdo, G.~Papadopoulos and P.~K.~Townsend,
  ``New supergravities with central charges and Killing spinors in (2+1)-dimensions,''
Nucl.\ Phys.\ B {\bf 467}, 183 (1996).
[hep-th/9505032].
}

\lref\PestunRZ{
  V.~Pestun,
  ``Localization of gauge theory on a four-sphere and supersymmetric Wilson loops,''
Commun.\ Math.\ Phys.\  {\bf 313}, 71 (2012).
[arXiv:0712.2824 [hep-th]].
}

\lref\KuzenkoRD{
  S.~M.~Kuzenko and G.~Tartaglino-Mazzucchelli,
  ``Three-dimensional N=2 (AdS) supergravity and associated supercurrents,''
JHEP {\bf 1112}, 052 (2011).
[arXiv:1109.0496 [hep-th]].
}

\lref\RomelsbergerEC{
  C.~Romelsberger,
  ``Calculating the Superconformal Index and Seiberg Duality,''
[arXiv:0707.3702 [hep-th]].
}

\lref\RomelsbergerEG{
  C.~Romelsberger,
  ``Counting chiral primaries in N = 1, d=4 superconformal field theories,''
Nucl.\ Phys.\ B {\bf 747}, 329 (2006).
[hep-th/0510060].
}

\lref\DolanQI{
  F.~A.~Dolan and H.~Osborn,
  ``Applications of the Superconformal Index for Protected Operators and q-Hypergeometric Identities to N=1 Dual Theories,''
Nucl.\ Phys.\ B {\bf 818}, 137 (2009).
[arXiv:0801.4947 [hep-th]].
}

\lref\ClossetRU{
  C.~Closset, T.~T.~Dumitrescu, G.~Festuccia and Z.~Komargodski,
  ``Supersymmetric Field Theories on Three-Manifolds,''
JHEP {\bf 1305}, 017 (2013).
[arXiv:1212.3388].
}

\lref\PeetersFQ{
  K.~Peeters, J.~Sonnenschein and M.~Zamaklar,
  ``Holographic decays of large-spin mesons,''
JHEP {\bf 0602}, 009 (2006).
[hep-th/0511044].
}

\lref\DubovskyTU{
  S.~Dubovsky, A.~Lawrence and M.~M.~Roberts,
  ``Axion monodromy in a model of holographic gluodynamics,''
JHEP {\bf 1202}, 053 (2012).
[arXiv:1105.3740 [hep-th]].
}

\lref\CasherWY{
  A.~Casher, H.~Neuberger and S.~Nussinov,
  ``Chromoelectric Flux Tube Model of Particle Production,''
Phys.\ Rev.\ D {\bf 20}, 179 (1979).
}

\lref\mb{
  G.~E.~Brown and M.~Rho,
  ``The multifaceted skyrmion.''
}

\lref\ZahedQZ{
  I.~Zahed and G.~E.~Brown,
  ``The Skyrme Model,''
Phys.\ Rept.\  {\bf 142}, 1 (1986).
}

\lref\KSii{
  K.~Kodaira and D.C.~Spencer,
  ``On Deformations of Complex Analytic Structures II,''
Ann. Math. {\bf 67}, 403 (1958).
}

\lref\Kodairabook{
K.~Kodaira, ``Complex Manifolds and Deformation of Complex Structures,'' Springer (1986).
}

\lref\Kobayashi{
S.~Kobayashi, ``Differential Geometry of Complex Vector Bundles,'' Princeton University Press (1987).
}

\lref\WittenDF{
  E.~Witten,
  ``Constraints on Supersymmetry Breaking,''
Nucl.\ Phys.\ B {\bf 202}, 253 (1982).
}

\lref\spivak{
M.~Spivak, ``Calculus on Manifolds,'' Perseus (1965).
}

\lref\ImamuraWG{
  Y.~Imamura and D.~Yokoyama,
  ``N=2 supersymmetric theories on squashed three-sphere,''
Phys.\ Rev.\ D {\bf 85}, 025015 (2012).
[arXiv:1109.4734 [hep-th]].
}

\lref\MartelliAQA{
  D.~Martelli and A.~Passias,
  ``The gravity dual of supersymmetric gauge theories on a two-parameter deformed three-sphere,''
[arXiv:1306.3893 [hep-th]].
}

\lref\AharonyDHA{
  O.~Aharony, S.~S.~Razamat, N.~Seiberg and B.~Willett,
  ``3d dualities from 4d dualities,''
JHEP {\bf 1307}, 149 (2013).
[arXiv:1305.3924 [hep-th]].
}

\lref\SpiridonovZA{
  V.~P.~Spiridonov and G.~S.~Vartanov,
  ``Elliptic Hypergeometry of Supersymmetric Dualities,''
Commun.\ Math.\ Phys.\  {\bf 304}, 797 (2011).
[arXiv:0910.5944 [hep-th]].
}

\lref\KinneyEJ{
  J.~Kinney, J.~M.~Maldacena, S.~Minwalla and S.~Raju,
  ``An Index for 4 dimensional super conformal theories,''
Commun.\ Math.\ Phys.\  {\bf 275}, 209 (2007).
[hep-th/0510251].
}

\lref\SenPH{
  D.~Sen,
  ``Supersymmetry In The Space-time $\R \times S^3$,''
Nucl.\ Phys.\ B {\bf 284}, 201 (1987).
}

\lref\Kodairasone{
  K.~Kodaira, ``Complex structures on $S^{1}\times S^{3}$,''
Proceedings of the National Academy of Sciences of the United States of America {\bf 55}, 240 (1966).
}

\lref\Belgun{
F.~A.~Belgun, ``On the metric structure of non-K\"ahler complex surfaces,''
Math. Ann. {\bf 317}, 1 (2000).
}

\lref\gaudorn{
P.~Gauduchon and~L.~Ornea, ``Locally conformally K\"ahler metrics on Hopf surfaces,'' Ann. Inst. Fourier {\bf 48}, 4 (1998).
}

\lref\AharonyHDA{
  O.~Aharony, N.~Seiberg and Y.~Tachikawa,
  ``Reading between the lines of four-dimensional gauge theories,''
JHEP {\bf 1308}, 115 (2013).
[arXiv:1305.0318 [hep-th]].
}

\lref\CassaniDBA{
  D.~Cassani and D.~Martelli,
  ``Supersymmetry on curved spaces and superconformal anomalies,''
[arXiv:1307.6567 [hep-th]].
}

\lref\JohansenAW{
  A.~Johansen,
  ``Twisting of $N=1$ SUSY gauge theories and heterotic topological theories,''
Int.\ J.\ Mod.\ Phys.\ A {\bf 10}, 4325 (1995).
[hep-th/9403017].
}

\lref\Spiridonov{
  V.~Spiridonov,
  ``Elliptic Hypergeometric Functions,''
[arXiv:0704.3099].
}

\lref\GabadadzeVW{
  G.~Gabadadze and M.~A.~Shifman,
  ``Vacuum structure and the axion walls in gluodynamics and QCD with light quarks,''
Phys.\ Rev.\ D {\bf 62}, 114003 (2000).
[hep-ph/0007345].
}

\lref\Rains{
  E.~M.~Rains,
  ``Transformations of Elliptic Hypergeometric Integrals,''
[math/0309252].
}

\lref\Bult{
  F.~van de Bult,
  ``Hyperbolic Hypergeometric Functions,''
University of Amsterdam, Ph.D. Thesis, [http://www.its.caltech.edu/$\sim$vdbult/Thesis.pdf].
}

\lref\KodairaCCASii{
K.~Kodaira, ``On the structure of compact complex analytic surfaces, II,'' American Journal of Mathematics {\bf 88}, 682 (1966).
}

\lref\SpiridonovHF{
  V.~P.~Spiridonov and G.~S.~Vartanov,
  ``Elliptic hypergeometry of supersymmetric dualities II. Orthogonal groups, knots, and vortices,''
[arXiv:1107.5788 [hep-th]].
}

\lref\BeniniNC{
  F.~Benini, T.~Nishioka and M.~Yamazaki,
  ``4d Index to 3d Index and 2d TQFT,''
Phys.\ Rev.\ D {\bf 86}, 065015 (2012).
[arXiv:1109.0283 [hep-th]].
}

\lref\RazamatOPA{
  S.~S.~Razamat and B.~Willett,
  ``Global Properties of Supersymmetric Theories and the Lens Space,''
[arXiv:1307.4381 [hep-th]].
}

\lref\Nakagawa{
N.~Nakagawa, ``Complex structures on $L (p, q)\times S^1$,'' Hiroshima Mathematical Journal {\bf 25} 423 (1995).
}

\lref\GaddeDDA{
  A.~Gadde and S.~Gukov,
  ``2d Index and Surface operators,''
[arXiv:1305.0266 [hep-th]].
}

\lref\ElitzurNR{
  S.~Elitzur, G.~W.~Moore, A.~Schwimmer and N.~Seiberg,
  ``Remarks on the Canonical Quantization of the Chern-Simons-Witten Theory,''
Nucl.\ Phys.\ B {\bf 326}, 108 (1989).
}

\lref\SonFH{
  D.~T.~Son, M.~A.~Stephanov and A.~R.~Zhitnitsky,
  ``Domain walls of high density QCD,''
Phys.\ Rev.\ Lett.\  {\bf 86}, 3955 (2001).
[hep-ph/0012041].
}
\lref\ForbesET{
  M.~M.~Forbes and A.~R.~Zhitnitsky,
  ``Domain walls in QCD,''
JHEP {\bf 0110}, 013 (2001).
[hep-ph/0008315].
}

\lref\BeniniNDA{
  F.~Benini, R.~Eager, K.~Hori and Y.~Tachikawa,
  ``Elliptic genera of two-dimensional N=2 gauge theories with rank-one gauge groups,''
[arXiv:1305.0533 [hep-th]].
}

\lref\BeniniXPA{
  F.~Benini, R.~Eager, K.~Hori and Y.~Tachikawa,
  ``Elliptic genera of 2d N=2 gauge theories,''
[arXiv:1308.4896 [hep-th]].
}

\lref\KapustinKZ{
  A.~Kapustin, B.~Willett and I.~Yaakov,
  ``Exact Results for Wilson Loops in Superconformal Chern-Simons Theories with
  Matter,''
  JHEP {\bf 1003}, 089 (2010)
  [arXiv:0909.4559 [hep-th]].
}

\lref\JafferisUN{
  D.~L.~Jafferis,
  ``The Exact Superconformal R-Symmetry Extremizes Z,''
JHEP {\bf 1205}, 159 (2012).
[arXiv:1012.3210 [hep-th]].
}

\lref\CecottiSA{
  S.~Cecotti,
  ``Higher Derivative Supergravity Is Equivalent To Standard Supergravity Coupled To Matter. 1.,''
Phys.\ Lett.\ B {\bf 190}, 86 (1987)..
}

\lref\HamaAV{
  N.~Hama, K.~Hosomichi and S.~Lee,
  ``Notes on SUSY Gauge Theories on Three-Sphere,''
JHEP {\bf 1103}, 127 (2011).
[arXiv:1012.3512 [hep-th]].
}

\lref\WittenKH{
  E.~Witten,
  ``Baryons in the 1/n Expansion,''
Nucl.\ Phys.\ B {\bf 160}, 57 (1979).
}

\lref\HamaEA{
  N.~Hama, K.~Hosomichi and S.~Lee,
  ``SUSY Gauge Theories on Squashed Three-Spheres,''
JHEP {\bf 1105}, 014 (2011).
[arXiv:1102.4716 [hep-th]].
}

\lref\AldayLBA{
  L.~F.~Alday, D.~Martelli, P.~Richmond and J.~Sparks,
  ``Localization on Three-Manifolds,''
[arXiv:1307.6848 [hep-th]].
}

\lref\MaNPF{
  Y.~L.~Ma and M.~Harada,
  ``Lecture notes on the Skyrme model,''
[arXiv:1604.04850 [hep-ph]].
}

\lref\NianQWA{
  J.~Nian,
  ``Localization of Supersymmetric Chern-Simons-Matter Theory on a Squashed $S^3$ with $SU(2)\times U(1)$ Isometry,''
[arXiv:1309.3266 [hep-th]].
}

\lref\KlareDKA{
  C.~Klare and A.~Zaffaroni,
  ``Extended Supersymmetry on Curved Spaces,''
JHEP {\bf 1310}, 218 (2013).
[arXiv:1308.1102 [hep-th]].
}

\lref\Toine{
  A. Van Proeyen
  ``${\cal N}=2$ Supergravity in $d=4,5,6$ and its Matter Couplings.''
http://itf.fys.kuleuven.be/$\sim$toine/LectParis.pdf
}

\lref\BhattacharyaZY{
  J.~Bhattacharya, S.~Bhattacharyya, S.~Minwalla and S.~Raju,
  ``Indices for Superconformal Field Theories in 3,5 and 6 Dimensions,''
JHEP {\bf 0802}, 064 (2008).
[arXiv:0801.1435 [hep-th]].
}

\lref\KapustinJM{
  A.~Kapustin and B.~Willett,
  ``Generalized Superconformal Index for Three Dimensional Field Theories,''
[arXiv:1106.2484 [hep-th]].
}

\lref\DimoftePY{
  T.~Dimofte, D.~Gaiotto and S.~Gukov,
  ``3-Manifolds and 3d Indices,''
[arXiv:1112.5179 [hep-th]].
}

\lref\BeniniUI{
  F.~Benini and S.~Cremonesi,
  ``Partition functions of N=(2,2) gauge theories on $S^2$ and vortices,''
[arXiv:1206.2356 [hep-th]].
}

\lref\DoroudXW{
  N.~Doroud, J.~Gomis, B.~Le Floch and S.~Lee,
  ``Exact Results in D=2 Supersymmetric Gauge Theories,''
JHEP {\bf 1305}, 093 (2013).
[arXiv:1206.2606 [hep-th]].
}

\lref\DoroudPKA{
  N.~Doroud and J.~Gomis,
[arXiv:1309.2305 [hep-th]].
}

\lref\CecottiQE{
  S.~Cecotti, S.~Ferrara, M.~Porrati and S.~Sabharwal,
  ``New Minimal Higher Derivative Supergravity Coupled To Matter,''
Nucl.\ Phys.\ B {\bf 306}, 160 (1988)..
}

\lref\KlareGN{
  C.~Klare, A.~Tomasiello and A.~Zaffaroni,
  ``Supersymmetry on Curved Spaces and Holography,''
JHEP {\bf 1208}, 061 (2012).
[arXiv:1205.1062 [hep-th]].
}

\lref\KutasovXB{
  D.~Kutasov,
  ``Geometry On The Space Of Conformal Field Theories And Contact Terms,''
Phys.\ Lett.\ B {\bf 220}, 153 (1989)..
}

\lref\SeibergPF{
  N.~Seiberg,
  ``Observations on the Moduli Space of Superconformal Field Theories,''
Nucl.\ Phys.\ B {\bf 303}, 286 (1988)..
}

\lref\CecottiME{
  S.~Cecotti and C.~Vafa,
  ``Topological antitopological fusion,''
Nucl.\ Phys.\ B {\bf 367}, 359 (1991)..
}

\lref\NishiokaHAA{
  T.~Nishioka and I.~Yaakov,
  ``Supersymmetric Renyi Entropy,''
[arXiv:1306.2958 [hep-th]].
}

\lref\ClossetVRA{
  C.~Closset, T.~T.~Dumitrescu, G.~Festuccia and Z.~Komargodski,
  ``The Geometry of Supersymmetric Partition Functions,''
[arXiv:1309.5876 [hep-th]].
}

\lref\deWitZA{
  B.~de Wit, S.~Katmadas and M.~van Zalk,
  ``New supersymmetric higher-derivative couplings: Full N=2 superspace does not count!,''
JHEP {\bf 1101}, 007 (2011).
[arXiv:1010.2150 [hep-th]].
}

\lref\HasenfratzGR{
  P.~Hasenfratz and G.~'t Hooft,
  ``A Fermion-Boson Puzzle in a Gauge Theory,''
Phys.\ Rev.\ Lett.\  {\bf 36}, 1119 (1976).
}

\lref\GoldhaberDP{
  A.~S.~Goldhaber,
  ``Spin and Statistics Connection for Charge-Monopole Composites,''
Phys.\ Rev.\ Lett.\  {\bf 36}, 1122 (1976).
}

\lref\Hijazi{
O. Hijazi, 
``A conformal lower bound for the smallest eigenvalue of the Dirac operator and Killing spinors,'' Communications in Mathematical Physics 104 (1986), no. 1, 151Ð162.
}

\lref\deRooMM{
  M.~de Roo, J.~W.~van Holten, B.~de Wit and A.~Van Proeyen,
  ``Chiral Superfields in $N=2$ Supergravity,''
Nucl.\ Phys.\ B {\bf 173}, 175 (1980).
}

\lref\OsbornGM{
  H.~Osborn,
  ``Weyl consistency conditions and a local renormalization group equation for general renormalizable field theories,''
Nucl.\ Phys.\ B {\bf 363}, 486 (1991)..
}

\lref\SekiMU{
  S.~Seki and J.~Sonnenschein,
  ``Comments on Baryons in Holographic QCD,''
JHEP {\bf 0901}, 053 (2009).
[arXiv:0810.1633 [hep-th]].
}

\lref\Huybrechts{
D.~Huybrechts, ``Complex Geometry: An Introduction,'' Springer (2006).
}

\lref\TanakaDCA{
  A.~Tanaka,
  ``Localization on round sphere revisited,''
[arXiv:1309.4992 [hep-th]].
}

\lref\AsninXX{
  V.~Asnin,
  ``On metric geometry of conformal moduli spaces of four-dimensional superconformal theories,''
JHEP {\bf 1009}, 012 (2010).
[arXiv:0912.2529 [hep-th]].
}

\lref\inprogress{
 O.~Aharony, F.~Bigazzi, A.~Cotrone, Z.~Komargodski, J.~Sonnenschein, in progress.
 }

\lref\DoroudXW{
  N.~Doroud, J.~Gomis, B.~Le Floch and S.~Lee,
  ``Exact Results in $D=2$ Supersymmetric Gauge Theories,''
JHEP {\bf 1305}, 093 (2013).
[arXiv:1206.2606 [hep-th]].
}

\lref\GomisWY{
  J.~Gomis and S.~Lee,
  ``Exact K\"ahler Potential from Gauge Theory and Mirror Symmetry,''
JHEP {\bf 1304}, 019 (2013).
[arXiv:1210.6022 [hep-th]].
}

\lref\ClossetVP{
  C.~Closset, T.~T.~Dumitrescu, G.~Festuccia, Z.~Komargodski and N.~Seiberg,
  ``Comments on Chern-Simons Contact Terms in Three Dimensions,''
JHEP {\bf 1209}, 091 (2012).
[arXiv:1206.5218 [hep-th]].
}

\lref\BeniniUI{
  F.~Benini and S.~Cremonesi,
  ``Partition functions of ${\cal N}=(2,2)$ gauge theories on $S^2$ and vortices,''
[arXiv:1206.2356 [hep-th]].
}

\lref\tHooftALW{
  G.~'t Hooft,
  ``A Planar Diagram Theory for Strong Interactions,''
Nucl.\ Phys.\ B {\bf 72}, 461 (1974).
}

\lref\GaiottoKFA{
  D.~Gaiotto, A.~Kapustin, N.~Seiberg and B.~Willett,
  ``Generalized Global Symmetries,''
JHEP {\bf 1502}, 172 (2015).
[arXiv:1412.5148 [hep-th]].
}

\lref\ClossetVP{
  C.~Closset, T.~T.~Dumitrescu, G.~Festuccia, Z.~Komargodski and N.~Seiberg,
  ``Comments on Chern-Simons Contact Terms in Three Dimensions,''
JHEP {\bf 1209}, 091 (2012).
[arXiv:1206.5218 [hep-th]].
}
\lref\ClossetVG{
  C.~Closset, T.~T.~Dumitrescu, G.~Festuccia, Z.~Komargodski and N.~Seiberg,
  ``Contact Terms, Unitarity, and F-Maximization in Three-Dimensional Superconformal Theories,''
JHEP {\bf 1210}, 053 (2012).
[arXiv:1205.4142 [hep-th]].
}

\lref\ClossetSXA{
  C.~Closset and I.~Shamir,
  ``The $\CN=1$ Chiral Multiplet on $T^2\times S^2$ and Supersymmetric Localization,''
[arXiv:1311.2430 [hep-th]].
}

\lref\WittenVV{
  E.~Witten,
  ``Current Algebra Theorems for the U(1) Goldstone Boson,''
Nucl.\ Phys.\ B {\bf 156}, 269 (1979).
}

\lref\VenezianoEC{
  G.~Veneziano,
  ``U(1) Without Instantons,''
Nucl.\ Phys.\ B {\bf 159}, 213 (1979).
}

\lref\AharonyMJS{
  O.~Aharony,
  ``Baryons, monopoles and dualities in Chern-Simons-matter theories,''
JHEP {\bf 1602}, 093 (2016).
[arXiv:1512.00161 [hep-th]].
}

\lref\BeniniAED{
  F.~Benini,
  ``Three-dimensional dualities with bosons and fermions,''
JHEP {\bf 1802}, 068 (2018).
[arXiv:1712.00020 [hep-th]].
}

\lref\HsinBLU{
  P.~S.~Hsin and N.~Seiberg,
  ``Level/rank Duality and Chern-Simons-Matter Theories,''
JHEP {\bf 1609}, 095 (2016).
[arXiv:1607.07457 [hep-th]].
}

\lref\GabadadzePP{
  G.~Gabadadze and M.~A.~Shifman,
  ``D walls and junctions in supersymmetric gluodynamics in the large N limit suggest the existence of heavy hadrons,''
Phys.\ Rev.\ D {\bf 61}, 075014 (2000).
[hep-th/9910050].
}

\lref\GaiottoTNE{
  D.~Gaiotto, Z.~Komargodski and N.~Seiberg,
  ``Time-reversal breaking in QCD$_{4}$, walls, and dualities in 2 + 1 dimensions,''
JHEP {\bf 1801}, 110 (2018).
[arXiv:1708.06806 [hep-th]].
}
\lref\GaiottoYUP{
  D.~Gaiotto, A.~Kapustin, Z.~Komargodski and N.~Seiberg,
  ``Theta, Time Reversal, and Temperature,''
JHEP {\bf 1705}, 091 (2017).
[arXiv:1703.00501 [hep-th]].
}

\lref\HsinVCG{
  P.~S.~Hsin, H.~T.~Lam and N.~Seiberg,
  ``Comments on One-Form Global Symmetries and Their Gauging in 3d and 4d,''
[arXiv:1812.04716 [hep-th]].
}

\lref\BeniniMF{
  F.~Benini, C.~Closset and S.~Cremonesi,
  ``Comments on 3d Seiberg-like dualities,''
JHEP {\bf 1110}, 075 (2011).
[arXiv:1108.5373 [hep-th]].
}

\lref\MooreYH{
  G.~W.~Moore and N.~Seiberg,
  ``Taming the Conformal Zoo,''
Phys.\ Lett.\ B {\bf 220}, 422 (1989).
}

\lref\WittenTX{
  E.~Witten,
  ``Current Algebra, Baryons, and Quark Confinement,''
Nucl.\ Phys.\ B {\bf 223}, 433 (1983).
}

\lref\AharonyDA{
  O.~Aharony, J.~Sonnenschein and S.~Yankielowicz,
  ``A Holographic model of deconfinement and chiral symmetry restoration,''
Annals Phys.\  {\bf 322}, 1420 (2007).
[hep-th/0604161].
}

\lref\NayaKYI{
  C.~Naya and P.~Sutcliffe,
  ``Skyrmions and clustering in light nuclei,''
[arXiv:1811.02064 [hep-th]].
}

\lref\IseHopfsurf{
M.~Ise,
   ``On the geometry of Hopf manifolds,''
  Osaka Math. J., 12, 1960, p.387--402.
}

\lref\PolchinskiUF{
  J.~Polchinski and M.~J.~Strassler,
  ``The String dual of a confining four-dimensional gauge theory,''
[hep-th/0003136].
}

\lref\PolyakovRS{
  A.~M.~Polyakov,
  ``Compact Gauge Fields and the Infrared Catastrophe,''
Phys.\ Lett.\ B {\bf 59}, 82 (1975), [Phys.\ Lett.\  {\bf 59B}, 82 (1975)].
}

\lref\TongKPV{
  D.~Tong,
  ``Lectures on the Quantum Hall Effect,''
[arXiv:1606.06687 [hep-th]].
}

\lref\GabadadzeFF{
  G.~Gabadadze and M.~Shifman,
  ``QCD vacuum and axions: What's happening?,''
Int.\ J.\ Mod.\ Phys.\ A {\bf 17}, 3689 (2002).
[hep-ph/0206123].
}

\lref\KomargodskiSMK{
  Z.~Komargodski, T.~Sulejmanpasic and M.~\"Unsal,
  ``Walls, anomalies, and deconfinement in quantum antiferromagnets,''
Phys.\ Rev.\ B {\bf 97}, no. 5, 054418 (2018).
[arXiv:1706.05731 [cond-mat.str-el]].
}

\lref\AdkinsYA{
  G.~S.~Adkins, C.~R.~Nappi and E.~Witten,
  ``Static Properties of Nucleons in the Skyrme Model,''
Nucl.\ Phys.\ B {\bf 228}, 552 (1983).
}

\lref\ArgurioUUP{
  R.~Argurio, M.~Bertolini, F.~Bigazzi, A.~L.~Cotrone and P.~Niro,
  ``QCD domain walls, Chern-Simons theories and holography,''
JHEP {\bf 1809}, 090 (2018).
[arXiv:1806.08292 [hep-th]].
}

\lref\MallHopfsurf{
D.~Mall,
   ``The cohomology of line bundles on Hopf manifolds,''
  Osaka J. Math., 28, 1991, p.999--1015.
}



\Title{}
{
\vbox{
\centerline{Baryons as Quantum Hall Droplets}
}
}
\centerline{Zohar Komargodski} 
\vskip15pt

\centerline{ {\it Simons Center for Geometry and Physics, Stony Brook, New York, USA}}

\centerline{ {\it and Weizmann Institute of Science, Rehovot 76100, Israel}}

\vskip20pt

\centerline{\bf Abstract}
\noindent   

We revisit the problem of baryons in the large $N$ limit of Quantum Chromodynamics. A special case in which the theory of Skyrmions is inapplicable is one-flavor QCD, where there are no light pions to construct the baryon from. More generally, the description of baryons made out of predominantly one flavor within the Skyrmion model is unsatisfactory. We propose a model for such baryons, where the baryons are interpreted as quantum Hall droplets.  An important element in our construction is an extended, 2+1 dimensional, meta-stable configuration of the $\eta'$ particle. Baryon number is identified with a magnetic symmetry on the 2+1 dimensional sheet. If the sheet has a boundary, there are finite energy chiral excitations which carry baryon number. These chiral excitations are analogous to the electron in the fractional quantum Hall effect. Studying the chiral vertex operators we are able to determine the spin, isospin, and certain excitations of the droplet. In addition, balancing the tension of the droplet against the energy stored at the boundary we estimate the size and mass of the baryons. The mass, size, spin, isospin, and excitations that we find agree with phenomenological expectations. 

\Date{January 2019}

\newsec{Introduction}

In this note we consider some problems concerning baryons in Quantum Chromodynamics (QCD). 
We take the gauge group to be $SU(N)$ and the number of flavors to be $N_f$. Throughout this paper, the quarks are taken to be degenerate with mass $M$ and $\theta_{QCD}=0$. The mass $M$ is real and non-negative.

It has been known for several decades that in the large $N$ 't Hooft limit~\tHooftALW\  baryons should be viewed as solitons~\refs{\WittenKH,\WittenTX}. 
This is because the effective coupling constant in large $N$ theories is $1/N$ while the mass of baryons scales linearly with $N$, suggesting that it can be understood, in some sense, as a soliton (the mass of solitons typically scales like the inverse of the coupling constant).

While the theory of baryons as classical solitons is on firm footing at large $N$, many of the results can be successfully applied also at finite $N$ and compared with particle physics data. See, for instance,~\refs{\ZahedQZ,\mb,\MaNPF} for reviews and references. We will also review some of the most important ingredients below. 

We will now discuss both the macroscopic and the microscopic points of view on baryons:

\subsec{Macroscopic Considerations}
If the quarks are light (but not necessarily massless) and $N_f$ is sufficiently small then due to chiral symmetry breaking QCD is well described at low energies by (pseudo-) Nambu-Goldstone fields moving on the group manifold $SU(N_f)$. For $N_f\geq 2$ we have that $\pi_3(SU(N_f))=\Z$ and this leads to a conserved charge, which should be identified with baryon number.
At large $N$ the classical solutions which carry this charge are trustworthy. Indeed, the action on the group manifold is proportional to $N$ and hence one has a systematic expansion around these solutions in powers of $1/N$. These solutions are called Skyrmions and they are identified with baryons.
Indeed, the baryons at large $N$ are heavy objects, with mass scaling linearly with $N$.
The quantization of the collective coordinates of the Skyrmion leads to a tower of baryons with various quantum numbers under $SU(N_f)$ and various values of the spin.

An important property of these baryons is that their ``size'' does not scale with $N$. This is simply because the factor of $N$ is an overall factor in the action (to leading order) and hence the classical equations of motion are independent of it. While this leads to many simplifications, it makes the detailed properties of baryons hard to predict because the precise classical solution would depend on the higher-derivative terms in the chiral Lagrangian. Many properties of the baryon (especially the qualitative ones) are however independent of these higher-derivative corrections. 
One can fix some of the parameters of the higher derivative terms and study detailed phenomenological properties.

Here we would like to make a comment, which puts the discussion below in a general perspective.\foot{We thank E.~Witten for this comment.}
At large $N$, QCD can be viewed as a nearly-free theory of mesons and glueballs. One can then ask, where are the baryons? 
 One could try to identify the baryon charge with $\pi_3$ of the {\it entire} space of fields. It is therefore a welcome surprise that it is sometimes enough to consider $\pi_3$ of the mesons that correspond to the (pseudo-) Nambu-Goldstone bosons of chiral symmetry breaking. However it is not guaranteed that this prescription correctly describes all the baryons.\foot{See, for instance,~\NayaKYI\ for a recent demonstration that enlarging the space of mesons may improve upon the predictions of the Skyrme model.} In fact, this is the main point of this note. We will consider some situations where the baryons are clearly not captured by just the (pseudo-) Nambu-Goldstone bosons. 
 
 The baryons we study here can be understood from the low-energy degrees of freedom, but these are not just the (pseudo-) Nambu-Goldstone bosons. This leads to a compelling new universal picture for those baryons.

To understand why it is necessary to go beyond the standard formalism associating the baryon charge with $\pi_3$ of the (pseudo-) Nambu-Goldstone bosons it is useful to consider the ``extreme'' case of $N_f=1$. One may think that since the $SU(N_f)$ group manifold trivializes there is no hope to have a macroscopic description of the baryons in that theory. However, there is one key point that we have neglected, which is that at large $N$ there is a light $\eta'$ particle if the quark's mass is parametrically below the QCD scale~\refs{\WittenVV,\VenezianoEC}.  The $\eta'$ particle leads to the $U(1)$ group manifold (perhaps with some potential on the circle). It may still seem hopeless that we can describe baryons macroscopically with just the $\eta'$ since $\pi_3(U(1))=0$.\foot{Of course, the same difficulty with describing baryons for $N_f=1$ arises in holography, see for instance~\SekiMU\ for a discussion.} But, as we will summarize below, it proves possible to give a macroscopic description of baryons in spite of the fact that $\pi_3(U(1))=0$. We will argue that some extended excitations of the $\eta'$ field carry a nontrivial Topological Field Theory (TFT) that arises due to the dynamics of some heavy (glueball) fields that re-arrange. The heavy fields that re-arrange are not visible at low energies, except that they lead to a subtle singularity on the $\eta'$ circle. In this sense, our construction involves the $\eta'$ (pseudo-) Nambu Goldstone as well as a subtle effect due to glueballs which lead to a TFT on some extended excitations of the $\eta'$ field. It is the combination of the $\eta'$ field and the TFT that allows us to define baryon number. Using these observations one can write explicit classical configurations which carry baryon number. These configurations are chiral density waves on the boundary of the sheet. We will explain this in detail.

Our description uncannily captures many qualitative properties of the baryon in one-flavor QCD correctly. In particular, we find that the typical size of the baryon does not scale with $N$, the mass scales linearly with $N$ and the the spin is $N/2$, exactly as expected from the microscopic description (see below).
We also generalize our description to $N_f>1$. There, in addition to the ordinary Skyrmions, we can consider our extended excitations of the $\eta'$ field and the accompanying TFT that lives on these extended objects. This again leads to objects with baryon number. We show that their isospin and spin exactly match what we expect from the phenomenology of baryons. Their mass, size, and excitations are likewise in agreement with general expectations. 

In our picture, the baryons correspond to vertex operators in a certain chiral algebra that lives on the boundary of a 2+1 dimensional sheet with a TFT. While both our description and the ordinary Skyrmions capture the quantum numbers correctly, as we will argue, it is natural to assume that for low-spin baryons the ordinary Skyrmions are more adequate and for high-spin baryons the new picture where baryons are vertex operators in a chiral algebra is more appropriate. Indeed, for $N_f=1$ the baryons are necessarily of high spin (since the wave function has to be symmetric in the spin quantum numbers -- see below) and the new picture involving the $\eta'$ and the TFT is the only available macroscopic description of such baryons.

\subsec{Microscopic Considerations}

When the quarks are very heavy (compared to the QCD scale) we are in the non-relativistic regime and hence the quarks do not move much. For this reason the spin essentially decouples from the dynamics. In particular, baryons that are made of just one flavor (whether $N_f=1$ or $N_f>1$) have spin $N/2$. This is because the wave function is completely anti-symmetric in color and hence it must be completely symmetric in spin, which means that all the spins are aligned. The baryon is essentially spherically symmetric in shape and the wave function takes the form of a product over one-particle wave functions as explained in~\WittenKH.
In fact, in this limit of heavy quarks it does not really matter whether the baryon is made of several different quark species or not, as this would only affect the structure of the wave function in the spin-isospin space but the coordinate space wave function would still take the form of a product and 
and the baryon would be essentially spherical. Since the wave function factorizes into a product of single-particle wave functions, it follows that the baryon size does not scale with $N$, in agreement with the prediction of the macroscopic model.

Things get more interesting when we lower the mass of the constituent quarks. The main effect that comes to play is that the quarks become more mobile and hence the spin-orbit and spin-spin couplings becomes more important. As long as the total spin is $O(1)$ in the large $N$ limit, this effect is again not  important since it does not compete with the other contributors to the energy of the baryon (such as the potential and kinetic energies, which are of order $N$) and the picture of a spherical baryon continues to hold to a good approximation. As we lower the mass of the constituent quarks further these baryons should be eventually describable by the macroscopic theory based on the group manifold $SU(N_f)$, elaborated upon above. This is essentially the logical basis for the literature on Skyrmions. 

By contrast, for baryons with total spin $O(N)$, as we lower the mass and the constituent quarks become more mobile, we now cannot neglect the spin-orbit and spin-spin interactions. It is likely that it produces a configuration that is not rotationally symmetric. One-flavor QCD allows to focus our attention on this phenomenon because the baryon has spin $N/2$ in the large mass limit and hence as the mass is lowered we might indeed worry about the spin-orbit and spin-spin couplings. Such high-spin baryons of course exists also if $N_f>1$. They have a high spin as long as they are made out of predominantly one flavor. These high-spin baryons are possibly spatially deformed.  Our description of these baryons, associating them to some edge excitations on the boundary of a 2+1 dimensional pancake (which carries a TFT) therefore reflects some of these old expectations that these baryons should be spatially deformed.

\subsec{Summary of Results and Outline}

Let us first summarize our results for $N_f=1$,  which distills the issues with high-spin baryons and illuminates some general new phenomena that occur in the chiral Lagrangian. 

 If the quark's mass is small but nonzero we argue that the theory admits excitations that look like infinite sheets (i.e. co-dimension 1 configurations). This is puzzling at first sight because from the point of view of the infrared theory these sheets carry a conserved charge and hence cannot decay. But there are no such unbreakable sheets in the full QCD theory. 
These sheets could decay by producing vortices from the vacuum. Unfortunately the mass of these vortices is incalculable in the infrared theory (it formally diverges). This is why from the point of view of the infrared observer the sheets seem stable. 
But in the full theory the vortices cost finite energy and the sheets are allowed to decay. An important ingredient in our proposal is that these sheets are actually meta-stable in the full theory. Their decay is through the spontaneous creation of holes in the sheet, which can be estimated to lead to a decay probability that scale like $e^{-N}$. 

Therefore, at large $N$ we are dealing with some extended excitations of the $\eta'$ field which are very long lived. It therefore makes sense to study the theory of these excitations as well as the properties of the boundaries of these sheets.\foot{We would like to thank N.~Seiberg for many illuminating discussions on this topic.} 

It turns out that due to the re-arrangement of some heavy glueballs on these sheets, the sheets carry a nontrivial topological field theory. In fact, these sheets should be thought of as droplets of a fractional Hall state. It is an Abelian quantum Hall state for $N_f=1$. The sheets generically eventually decay. However, we show that the boundary of the sheet has a chiral algebra and some excitations in the chiral algebra carry baryon number. If the boundary has such an excitation then the sheet can no longer completely decay due to the conservation of baryon number.
This is quite analogous to the Skyrmions -- large Skyrmions always tend to shrink but we know that the baryon number that the carry will prevent them from disappearing altogether.  Interestingly, the baryon gauge field couples to the TFT on the sheet exactly in the same fashion that electromagnetism couples to the fractional quantum Hall state. 

The essential new point which allows us to make some progress on this old problem of baryons in the large $N$  limit  
is the TFT in the low-energy theory on the sheet. This is the key ingredient which allows to write configurations on the sheet which carry baryon number. Indeed, the TFT degrees of freedom on the sheet allow to write explicit configurations with baryon number. This construction therefore bridges the gap between the ultraviolet baryons and the infrared effective theory;
We now have a description of the baryons in terms of objects in the infrared  field theory, where the effective coupling constant is small. The baryons appear as solitons in our effective field theory in a very precise sense: the theory on the edge of the sheet has a big circle in target space and the baryon is the heavy configuration which winds around it. 

This formalism allows us to compute robustly many properties of these baryons. 
 The baryons we find have mass $\sim N$, spin $N/2$, and their size that does not scale with $N$. Furthermore, small excitations of this baryon also behave as expected in terms of their large $N$ scaling. The theory of baryons therefore essentially boils down to studying the properties of Hall droplets. These droplets have some tension that tends to shrink the droplet but the baryon excitation at the edge counteracts this attractive force. For large enough droplets we can compute precisely these two effects and we can see that they counteract each other. 

 Similarly to the usual theory of Skyrmions, while many qualitative properties do not depend on the properties of various higher-derivative interactions, to make quantitative predictions (e.g. concerning the exact energy and charge distributions) one needs to invoke a concrete model of the  properties of the Hall droplet.  Indeed, while we can compute the tension of the droplet and the force due to the edge mode reliably for large droplets, these two effects offset each other when the size of the droplet is comparable to the strong interactions scale. This is essentially good news because we expect the baryon to be of that size.
But that means that, as in all the Skyrmion models, to make quantitative predictions we are required to make some assumptions about what happens in that regime of small droplets.  

An interesting general phenomenological prediction of what we find is that the low lying excitations of these baryons are chiral modes. This is entirely not obvious from the microscopic point of view. Since the baryon should be thought of as a Hall droplet, at least in some naive sense, it is flattened into a pancake-like shape so the fact that the excitations are chiral is certainly possible. Since the baryon has size of order the inverse strong coupling scale, the energy gap to these chiral excitations is of order the strong coupling scale, so in practice it may not be easy to disentangle them from other excitations. (Since the sheet is long lived one can also imagine, as a thought experiment, pumping energy to the system and stretching the droplet. Then, the gap to exciting the chiral edge modes shrinks and they can be then cleanly disentangled from 3+1 dimensional bulk excitations. In this thought experiment it is also obvious that the baryon density is supported near the edge of the droplet, namely, the quarks are pushed to the boundary of the droplet and arise as edge excitations.)

Another general surprising qualitative prediction concerns with the glue content of the baryon. While for simple (low-spin) baryons we 
expect that the quarks will be connected by flux tubes, the gluon content of this high-spin baryon is quite exotic: it is a pancake with a topological field theory. When we bring some external quarks nearby this pancake, they become de-confined and pick up fractional spin (so they become anyons on the baryon).\foot{A similar exotic state of the gluons arises in the study of domain walls for $\theta_{QCD}=\pi$ (see, for instance,~\refs{\GaiottoYUP,\KomargodskiSMK,\GaiottoTNE,\HsinVCG}). In that context, however, the domain wall cannot end because it separates two distinct vacua, and for the same reason, it cannot decay through the creation of holes.} Furthermore, one can invest some energy and bring one of the constituent quarks of the baryon into the bulk of the pancake; this also makes it behave like an anyon. 

Now let us briefly summarize our results for $N_f>1$. For $N_f>1$ there are two main new ingredients: 

\item{1.} In addition to baryon number, there is isospin symmetry $SU(N_f)$. The baryons (along with all the other states in the theory) transform under some representations of $SU(N_f)$. Consider the baryon operators $B_{(s_1,f_1),\cdots, (s_N,f_N)}=\epsilon_{a_1,...,a_N}\Psi^{a_1}_{s_1,f_1}\cdots \Psi^{a_N}_{s_N,f_N}$. Here $s_i\in\{1,2\}$ corresponding to the spin and $f_i\in\{1,...,N_f\}$ corresponding to the isospin quantum numbers.
Since it is completely anti-symmetric in the color indices it must be  symmetric under the exchange of any pair $(s_i,f_i)\leftrightarrow(s_j,f_j)$. Therefore, the baryon is in the completely symmetric representation
\eqn\symmrep{Sym^N\left({1\over 2}\otimes \square\right)~,}
where $\square$ stands for the fundamental representation of isospin.

\item{2.} In addition to the Hall droplet baryons, there are also the ordinary Skyrmions that originate from $\pi_3(SU(N_f))$.

\medskip

We identify the TFT on the extended sheet.  
It is now a non-Abelian Chern-Simons theory. An encouraging result is that our TFT admits natural boundary conditions so that the states of the droplet are in representations of isospin and baryon number. This allows us to compare these states with QCD. 
Considering the chiral edge modes, we find that the object which carries baryon number miraculously furnishes precisely the correct representation~\symmrep\ for the spin $N/2$ baryon, i.e. the representation $Sym^N\left(\square\right)$ under isospin. We discuss the relation to the ordinary Skyrmions and suggest that the Hall droplet description is more suitable for the high-spin baryons.
As before, the mass,  size, and excitations of the baryons are in qualitative agreement with general expectations, while and spin and isospin quantum numbers exactly agree with the phenomenological expectations. 

The outline of the note is as follows. In section 2 we discuss a model in 2+1 dimensions which fleshes out some useful analogies for our main discussion. In section 3 we present the main results for $N_f=1$ QCD. In section 4 we carry out the analysis for $N_f>1$. Our discussion of $N_f>1$ is quite brief; many issues are left for the future.

\newsec{Warm-Up: Confinement in 2+1 Dimensional $U(1)$ Gauge Theory } 
A pedagogically useful model to consider is $U(1)$ gauge theory in 2+1 dimensions (with no Chern-Simons term) with a charge 1 scalar, $\Phi$.
\eqn\UVLag{{\cal L} = {1\over 2e^2 } F^2+|D_a\Phi|+V(\Phi)~.}  We take the potential for the scalar $\Phi$ to be  
$$V(\Phi)=M^2|\Phi^2|+\lambda|\Phi^4|~,$$
and we assume that $M^2>0$ and very large compared to all the other energy scales in the problem. (The Lagrangian is written in Euclidean signature.) This allows us to safely integrate out $\Phi$. The infrared theory is described by a $U(1)$ gauge field, which can be dualized to a compact pseudo-scalar $\varphi$ with periodicity $2\pi$,
$$\varphi\simeq\varphi+2\pi~.$$ 
We will further assume that in the original gauge theory monopole operators with a small coefficient are added to the Lagrangian, such that in terms of the dual scalar $\varphi$ there is a potential. For simplicity, we will take the effective theory in the infrared to be 
\eqn\PolyakovM{{\cal{L}} = \half e^2( \del \varphi)^2-\epsilon^3\cos(\varphi)~,}
where we suppressed higher-derivative terms and various possible corrections to the potential. We will discuss them soon.
$e^2$ and $\epsilon$ have dimension of a mass and $\epsilon$ is taken to be positive. 
The theory describes a non-degenerate trivially gapped vacuum at $\varphi=0\ \mod\ 2\pi$. 

The theory~\PolyakovM\ admits a conserved current with two indices 
\eqn\spintwo{J_{\mu\nu} ={1\over 2\pi} \epsilon_{\mu\nu\rho} \del^\rho \varphi~.}
In the terminology of~\GaiottoKFA\ (and see references therein), we can refer to~\spintwo\ as a $U(1)$ one-form symmetry.
This current is conserved even if we further complicate the potential in~\PolyakovM\ or add arbitrary higher-derivative interactions. It is conserved simply because the space of $\varphi$ configurations is a circle and $\pi_1(S^1)=\Z$.
The charged objects are strings. On one side of the string $\varphi$ is at $0$ and on the other side it is at $2\pi$. Therefore, on both sides of the string we are in the same unique trivially gapped vacuum.\foot{Since the vacuum is the same on both sides of the string, we prefer not to refer to this string as a domain wall.} The string has finite tension and finite width. From the perspective of a deep low-energy observer, it is a co-dimension 1 object. The tension of the string scales like 
\eqn\tens{T\sim \epsilon^{3/2} e}
and the width of the string scales like 
$$D\sim \epsilon^{-3/2}e~.$$
The effective theory on the string consists, at low energies, of the center of mass degree of freedom, which describes small transverse fluctuations of the string. There are no other light fields on the worldvolume of the string. To leading order, the Lagrangian of the center of mass degree of freedom is given by the Nambu-Goto term.

\ifig\Nfone{A string that ends on vortices. }%
{\epsfxsize3.5in\epsfbox{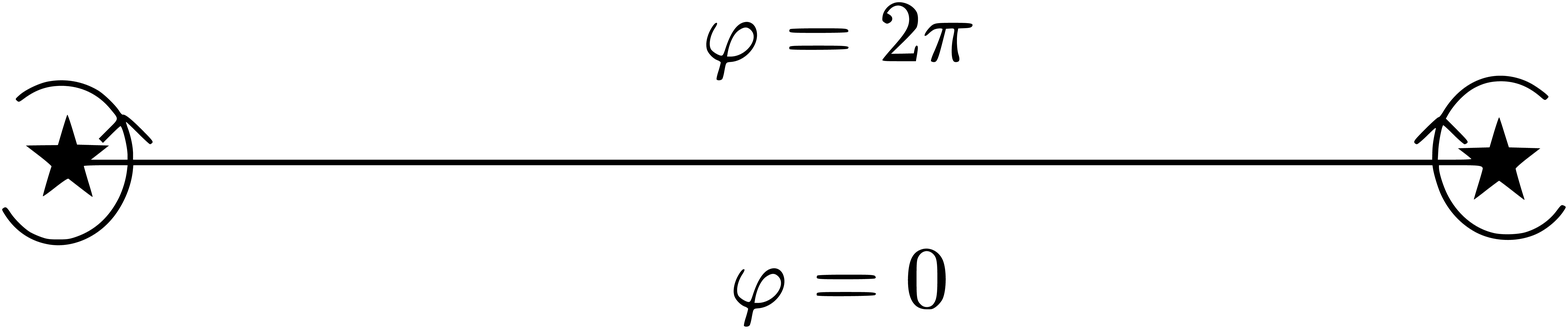}}

The conservation of~\spintwo\ guarantees that this string, once created, is unbreakable. It is useful to understand why it is unbreakable from the low-energy point of view. At first sight it seems possible to terminate the string with a vortex, namely, a point around which $\varphi$ has a monodromy $\varphi\to\varphi+2\pi$. See~\Nfone. However, the energy near this end-point would be bounded below by a constant times $\int d^2x {1\over r^2} (\del_\theta\varphi)^2 $ 
with $\theta$ the angle around the vortex and $r$ the radial coordinate. $\int d\theta (\del_\theta\varphi)^2$ is bounded below by ${1\over 2\pi}\left(\int d\theta \del_\theta\varphi\right)^2=2\pi$. Therefore $\int d^2x {1\over r^2} (\del_\theta\varphi)^2 $  is bounded below by 
 $2\pi\int  {dr\over r}  $, which diverges logarithmically in the ultraviolet (there is no infrared divergence if there are two vortices, such as in~\Nfone). This means that as far as the effective theory is concerned the mass of these vortices is incalculable.\foot{Since the effective theory breaks down close to the core of the vortex, the mass of these vortices is incalculable rather than infinite. Adding higher-derivative terms to~\PolyakovM\ would introduce power divergences to the mass of the vortex on top of the logarithmic divergence we are discussing.} 
 It is therefore impossible to create such vortices from the point of view of the effective theory and this is why the string charge is conserved. Whether these vortices ultimately have a finite mass depends on the ultraviolet of the theory and cannot be inferred from the $\varphi$ effective theory alone.

In our model~\UVLag\ there is no such conserved current~\spintwo\ and therefore we do expect that such vortices exist and have a finite mass, allowing to screen the string. Therefore, we should regard the symmetry associated to the current~\spintwo\ as an accidental one-form symmetry. (It is accidental though in a stronger sense than an ordinary accidental symmetry: here the symmetry remains regardless of the local operators made out of $\varphi$ that are added to~\PolyakovM.) More conceptually, the stability of the string from the infrared point of view is guaranteed by the fact that the field space is non-simply connected. In the full theory that circle is contractible (contracting it requires climbing up a potential barrier).

To understand how this symmetry is broken we need to identify the excitations in the full theory that regulate the divergence at the core of the vortex. In fact, it is easy to identify the vortices with the charged scalar $\Phi$ in our theory~\UVLag. The string is essentially the ``confining'' string in the phase with large positive $M^2$ and the vortices are the electric particles $\Phi$ which are ``confined.'' This is the confinement mechanism of~\PolyakovRS. Therefore, the mass of the vortex is indeed finite in the full theory, and it is related to the mass of the charged particle $M$. This identification makes perfect sense since the one-form symmetry is broken in the full theory by these charged particles $\Phi$. 

We could also imagine scenarios where, for example, the only dynamical particle has charge 2 under the $U(1)$ gauge symmetry in which case the fundamental string cannot be screened and the full theory has a $\Z_2$ symmetry protecting the fundamental string. This $\Z_2$ symmetry would be then embedded in the $U(1)$ symmetry~\spintwo. In this case, the elementary vortex truly has infinite mass and the string would be unbreakable. But two copies of the string could be broken by the creation of a double-vortex, which now has finite mass.

We could estimate the probability of decay per unit length per unit time in the full theory in the thin-string approximation (it is justified when the mass of $\Phi$, $M$, is large). For this we only need to know the end-point (vortex) mass and the tension~\tens. The action of creating a hole of length $L$ in the string consists of the contribution from the tension $TL^2$ and from the mass of the two end points, $ML$ (we neglect order one factors). The total Euclidean action of the instanton is thus  minimized for $L_0\sim M/T$, which is the typical size of the hole and the action is thus $S_0\sim M^2/T$, which leads to a decay probability, $\Gamma$, that behaves like 
$$\log \Gamma \sim - M^2/T~.$$
In particular, the string is very long lived for large $M$ and if the mass is infinite, the string is exactly stable. The length scale of the typical hole $L_0\sim M/T$ can be understood intuitively as follows: We can ask, given a string of length $L$, at what point does it pay off to create a pair of $\Phi$ particles from the vacuum and screen the string. The energy in the string is of order $TL$, while the energy of creating a pair from the vacuum scales like $M$. Therefore, it becomes advantageous to screen the string starting from the length scale $L_0$. For some original literature on the process of string snapping due to such instantons see~\CasherWY. See also~\PeetersFQ\ for the holographic perspective.  

\newsec{One-Flavor QCD at Large $N$}

We now study QCD in 3+1 dimensions with gauge group $SU(N)$ and with one Weyl fermion in the fundamental representation, $\Psi$, and one Weyl fermion in the anti-fundamental representation, $\tilde \Psi$. This is the usual one-flavor QCD model. We will include a mass term for the quarks $iM\Psi\tilde \Psi+c.c.$ and we will assume that $M$ is real and positive and that $\theta_{QCD}=0$.  If the mass $M$ is very large compared to the strong coupling scale then the baryon 
$$B_{s_1,...,s_N}=\epsilon_{a_1,...,a_N}\Psi^{a_1}_{s_1}\cdots \Psi^{a_N}_{s_N}~,$$ can be well described by treating the quarks non-relativistically. (The $s_i$ are the spin indices, taking the values 1 or 2. The baryon must be fully symmetric in these indices.)
Here we will be mostly interested in the opposite limit, where the mass $M$ is small compared to the strong coupling scale. Then, in the large $N$ limit, the long-distance theory consists of the $\eta'$ meson~\refs{\WittenVV,\VenezianoEC}. 

The $\eta'$ field is a periodic field, valued on the circle, as before, $\eta'\simeq\eta'+2\pi$. The effective theory is then given by 
\eqn\etapt{{\cal L} = F_\pi^2\left[\half (\del\eta')^2-\Lambda_{QCD}M\cos\eta'\right]~.}
Above, $\Lambda_{QCD}$ stands for the scale of strong interactions and $F_\pi$ is the decay constant, which has dimensions of mass. (We essentially define the scale of strong interactions $\Lambda_{QCD}$
such that the coefficient of the cosine above is just 1.) Since $F_\pi^2\sim N$ in the large $N$ limit, the above Lagrangian is of order $N$. There are important corrections to this Lagrangian at the next order in the $1/N$ expansion, but we will neglect them in the meanwhile. In addition, there are higher-derivative terms and various other terms that depend on higher orders in $M$. We will discuss soon the effects of all these.

Clearly the physics of the model~\etapt\ is in some respects quite analogous to that of~\PolyakovM, so some of the things we have studied in that context can be now carried over with only minor adaptations. In particular, $\eta'=0\ \mod\ 2\pi$ is the unique trivially gapped ground state of the theory.\foot{The $\eta'$ potential at the next order in the $1/N$ expansion is also minimized at $\eta'=0$, so that is indeed the unique ground state of the theory.}
The model has a topologically conserved current with three indices (this is a two-form $U(1)$ symmetry) associated with $\pi_1(S^1)$,
\eqn\threei{J_{\mu\nu\rho}={1\over 2\pi}\epsilon_{\mu\nu\rho\sigma}\del^\sigma \eta'~.}
The charged objects are now sheets (we will soon find a more suitable name for them), across which $\eta'$ changes from $0$ below the sheet to $2\pi$ above the sheet.

\ifig\Sheetpic{A Confining sheet that ends on a vortex loop. }%
{\epsfxsize3.5in\epsfbox{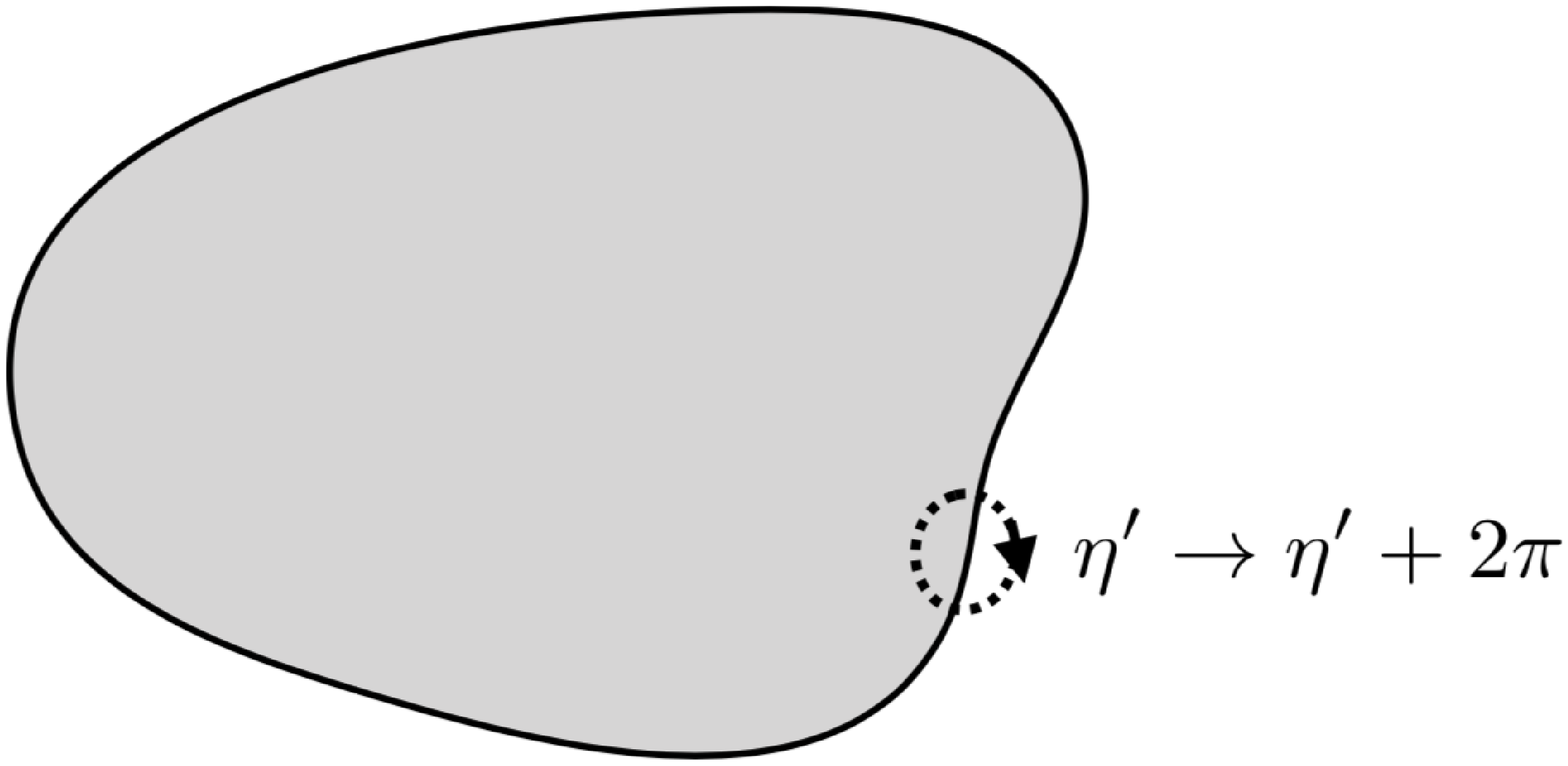}}

These sheets could end on some string-like objects (we may also refer to them as vortex loops -- they are the analogs of the charged particle in the previous section) around which $\eta'$ has the monodromy $\eta'\to \eta'+2\pi$. Within the effective Lagrangian~\etapt\ the tension of these strings is incalculable. It diverges logarithmically in the ultraviolet, as before. It is important to note that this logarithmic divergence now scales like 
\eqn\vortextension{T_v \sim F_\pi^2\int {dr\over r}\sim N\Lambda_{QCD}^2 \log \Lambda_{cutoff}~.}
($T_v$ denotes the tension of the vortex loop.)
Our main puzzle is to identify what are these objects in the full theory and what are their quantum numbers. One might think that it could be the confining string. But the tension of the confining string does not scale linearly with $N$ (the tension of the string is $O(N^0)$ in the large $N$ limit) while the tension of the vortex scales linearly in $N$, and furthermore, the confining string can be broken by creating $\Psi,\tilde \Psi$ pairs from the vacuum. In addition, the ordinary confining strings are not themselves confined by sheets. So, clearly, the vortex loops are not the ordinary confining strings.
The vortex loops are certain objects in the full theory which are confined by the sheets, in the same way that the vortex particles in the Polyakov model are confined by the strings. See~\Sheetpic.

Our main objective is to understand the effective theory on the sheet and then the allowed boundary conditions, which will shed some light on the nature of these vortex loops. 

Before we study the effective field theory on the sheet and its boundaries, one has to argue that the question makes sense. We anticipate that the divergence in the energy of the vortex loop will be regulated in the full theory, since the sheet must be able to decay in the full theory. So we need to estimate the lifetime of the sheet. Since the tension of the vortex loop scales like $N$, and since the tension of the sheet itself scales like $N$ (both follow simply from the fact that the Lagrangian~\etapt\ is of order $N$), it immediately follows that the decay probability per unit volume and per unit time, $\Gamma$, obeys 
\eqn\decayp{\log \Gamma \sim - N~.} 
While it is hard to compute the decay probability in detail due to the fact that the size of a typical hole is of the order of the QCD strong coupling scale, the $N$ scaling should be robust due to the fact that the action scales like $N$.\foot{This can be reproduced in the String Theory construction of such models, for instance in the setup of~\refs{\AharonyDA,\DubovskyTU,\ArgurioUUP}. Closely related issues can be also raised in the context of~\PolchinskiUF.} Since our extended 2+1 dimensional sheet  is meta-stable, the question of finding the effective theory on it makes sense. 

An intuitive way to understand the scaling~\decayp\ is to recall again that the conserved current~\threei\ is due to the $\pi_1$ of the $\eta'$ space of configurations. 
But it could be that in the full theory that circle is contractible if we climb up some potential barrier. Since $\eta'$ is the phase of the condensate $\langle\Psi\tilde \Psi\rangle$, the top of the potential barrier is a point where the condensate vanishes. The potential energy for the condensate scales linearly in $N$ since it comes from the disc diagram. Therefore, the potential energy barrier to un-winding $\eta'$ and hence poking a hole in the sheet scales linearly in $N$. See also a related analysis in~\refs{\ForbesET,\SonFH,\GabadadzeFF}. 

Normally, it would be possible to determine the effective theory on the sheet from the $\eta'$ Lagrangian alone~\etapt. But then there would no natural way to couple baryon number to any of these excitations. 
The point is that the $\eta'$ effective theory~\etapt\ is incomplete -- there is a cusp singularity which reflects an important effect missed by the $\eta'$ theory. It is because the $\eta'$ theory is incomplete that we can re-interpret some excitations of the theory as baryons.

Let us therefore  study the theory on the sheet more closely. From the point of view~\etapt\ the sheet is just a feature-less object through which $\eta'$ winds around the $S^1$. But here there is a crucial and conceptual difference from the 2+1 dimensional toy model for the confining string of section 2. In fact, the $S^1$ parameterized by the $\eta'$ field is not smooth. There is a cusp at $\eta'=\pi$. This cusp is not visible at the level of~\etapt\ but it appears clearly when we include $1/N$ corrections. Indeed, at the next order we find a term proportional to $$\min_{m\in \Z}{(\eta'+2\pi m)^2}~,$$ which is a perfectly smooth and $2\pi$ periodic function except at $\eta'=\pi\ \mod \ 2\pi$. Such a cusp singularity in an effective theory is quite mild. It does not signal the appearance of new light particles in 3+1 dimensions. 
The physical interpretation of the cusp is that some heavy fields need to rearrange at that point (that means that some heavy fields jump from one vacuum to another) and that is why the effective field theory description~\etapt\ fails. It is a rather subtle failure of effective field theory which has some consequences also in 3+1 dimensions, but perhaps its most dramatic effects arise when one studies the effective theory of the sheet.

The first important fact is that since some heavy fields rearrange at $\eta'=\pi$, the tension of the sheet is different from the result that would follow from just solving the appropriate equations of motion of~\etapt. Indeed, the contribution from just the motion of $\eta'$ across the sheet scales like 
$ T\sim F_\pi^2 M^{1/2}\Lambda_{QCD}^{1/2}\sim NM^{1/2}\Lambda_{QCD}^{5/2}$. But due to the rearrangement of the heavy fields the tension is in fact 
\eqn\tensionsheet{T\sim N\Lambda_{QCD}^3~.}
This scaling~\tensionsheet\ follows from general large $N$ arguments. It can be obtained also from the following (somewhat roundabout) considerations: Our sheet may decay through non-perturbative effects in the $1/N$ expansion~\decayp\ because it does not separate two distinct vacua and because it does not carry a conserved charge in the full theory. However, there is a closely related construction where a similar sheet is exactly stable since it separates two distinct stable vacua (then, creating a hole by a tunneling effect is inconsistent and the wall is stable). This situation was analyzed in detail in~\refs{\GabadadzeFF,\GaiottoTNE} (see also~\GaiottoYUP). One conclusion relevant for us here is that the contribution to the tension from the $\eta'$ field crossing the cusp must scale like~\tensionsheet. 

Before proceeding to studying the effects of this cusp on the effective theory on the sheet, we must make some cautionary remarks. Within an effective theory such as~\etapt, for sufficiently small $M$, the sheet is guaranteed to be a meta-stable configuration in the theory, decaying through non-perturbative effects that scale like~\decayp. However due to the cusp singularity at the core of the sheet, and the fact that some heavy fields are re-arranged at the core of the sheet, strictly speaking we cannot prove that the sheet must be a meta-stable configuration in the theory. It may disappear altogether. While one can give several qualitative arguments supporting the existence of the sheet in spite of the slight singularity in the effective theory, at present this is not proven and should be viewed as an assumption.\foot{This assumption holds true in holographic models, as explained in~\inprogress.}

The effective theory on the sheet always includes the center of mass degree of freedom, described by the Nambu-Goto action. This Nambu-Goto action captures small transverse fluctuations of the sheet. When we write below the effective theory of the sheet below we omit the Nambu-Goto part.
The most important effect from crossing the singularity at $\eta'=\pi$ can be again understood from the analysis of~\refs{\GaiottoTNE,\GaiottoYUP}. It turns out that the sheet acquires a Topological Field Theory (TFT) that lives on its surface. This TFT was identified in~\refs{\GaiottoTNE,\GaiottoYUP} with the $SU(N)_{-1}$ TFT: 
\eqn\TFTSUN{{-1\over 4\pi} \int_{{\cal M}_3} \Tr \left(\tilde a\wedge d\tilde a+{2\over 3}\tilde a^3\right)~,}
where ${\cal M}_3$ is the 2+1 dimensional space spanned by the sheet and $\tilde a$ is the $su(N)$-valued gauge field. 
Let us explain the heuristic reason for the appearance of this particular TFT. Consider the fermion condensate 
$\langle\Psi\tilde \Psi\rangle\sim N \Lambda^3 e^{i\eta'}~.$
If we perform an axial transformation, $\Psi\to e^{i\alpha}
\Psi,\tilde\Psi\to e^{i\alpha}\tilde \Psi$, then this leads to $\eta'\to \eta'+2\alpha$ accompanied by $\theta_{QCD}\to \theta_{QCD}+2\alpha$. Taking $\alpha=\pi$ (which is just fermion number symmetry) we see that $\theta_{QCD}$ returns to itself mod $2\pi$ and $\eta'$ shifts by $2\pi$. So the sheet in effect implements the transformation $\theta_{QCD}\to\theta_{QCD}+2\pi$. The Chern-Simons theory $SU(N)_{-1}$ therefore  arises in essence because on one side of the sheet we have $\theta_{QCD}=0$ and on the other side $\theta_{QCD}=2\pi$. (This is not a proof, it is simply a heuristic explanation.)

We now need to understand how to couple the baryon background gauge field $A^B$ to the TFT~\TFTSUN.
In fact, the most convenient way to do so is to first use level-rank duality:
\eqn\lrd{SU(N)_{-1}\longleftrightarrow U(1)_N~.}
We therefore introduce a $u(1)$ gauge field $a$ and write the theory on the sheet (again, we are omitting the Nambu-Goto sector associated to the fluctuations of the sheet in the orthogonal direction) as 
\eqn\sheetextended{S_{sheet}=\int_{{\cal M}_3} {N\over 4\pi} a\wedge da+{1\over 2\pi} a\wedge dA^B~.  }
Indeed, level-rank duality~\lrd\ exchanges the baryon symmetry in the $SU(N)_{-1}$ description with the magnetic symmetry in the description based on $U(1)_N$.\foot{An extensive discussion of the map between the baryon symmetry and the magnetic symmetry in level-rank duality can be found in~\HsinBLU.}  This allows us to write the coupling of the baryon gauge field to the low energy theory rather explicitly~\sheetextended. We see that our sheet is in fact a droplet realizing the fractional quantum hall effect. For this reason, we henceforth refer to the sheet as a ``Hall droplet.'' In fact, the coupling of $A^B$ above is precisely like the coupling of the electromagnetic gauge field to the emergent gauge field in the fractional quantum Hall effect.\foot{We should also determine the baryon gauge field counter-term in~\sheetextended, namely, the coefficient of the term $A^B\wedge dA^B$. Such a term does not affect the dynamics on the edge or the anyons in the bulk, but it is important in order to extract the Hall conductivity. Typically only the fractional part of the conductivity  is physically meaningful~\refs{\ClossetVG,\ClossetVP} but in the present situation the full theory is four dimensional, and no four-dimensional local terms can change the coefficient of  $A^B\wedge dA^B$ on the sheet. In other words, also the integer part of the Hall conductivity is meaningful in our case. We claim that in fact such a term is not present and hence the baryon current Hall conductivity is precisely $1/N$, as follows from~\sheetextended. A quick way to understand this is to revisit the argument below~\TFTSUN. Due to the mixed  axial-baryon anomaly, not only $\theta_{QCD}$ but also the theta term for the baryon background gauge field shifts under $\eta'\to\eta'+2\pi$. Since the quark carries baryon charge $1/N$, this leads to a shift of the baryon theta angle by $2\pi/N$, which is exactly reproduced by the $1/N$ conductivity on the sheet.  Another comment is that in this paper we focus on the implications of~\sheetextended\ for baryons at zero temperature and zero baryon chemical potential. However, this Chern-Simons theory on the sheet~\sheetextended\ could be also important for other questions, concerning various phenomena at finite baryon density. We leave this topic for the future.
}

It is now straightforward to study the effects of having a boundary for ${\cal M}_3$. (See~\refs{\MooreYH,\ElitzurNR} for the general analysis and~\TongKPV\ for a review.) 
We will assume that the boundary is a circle of radius $L$. There are various possible boundary conditions and boundary degrees of freedom. They are constrained by consistency with the bulk being a Chern-Simons theory. The most natural choice (and the one with the smallest possible central charge) is given in terms of the description~\sheetextended. We simply impose Dirichlet boundary conditions for the gauge field $a$.\foot{Let us for a moment take the boundary to be a straight line along the $y$ direction. The boundary term in the variation of the Chern-Simons action~\sheetextended\ is proportional to $a_t\delta a_y-a_y\delta a_t$. This can be made to vanish by the general Dirichlet boundary condition $a_t=va_y$ where $v$ is the velocity parameter; it is the velocity of the chiral edge mode. In the present context, we have relativistic invariance in the (y,t) plane and hence we must have the edge mode moving at the speed of light. In some of the literature on the Chern-Simons theory with boundary, the boundary condition $a_t=0$ is used instead. Hence, the reader must be careful when comparing results with the literature. We thank O.~Aharony for a discussion of this point.} This implies that the allowed gauge transformations become trivial on the boundary. Hence, the Hilbert space of states on the droplet is in representations of a global $U(1)$ symmetry. 
This $U(1)$ symmetry (which is the remnant of gauge transformations on the boundary) is precisely the same as the baryon symmetry $U(1)_B$. This identification follows from the coupling ${1\over 2\pi}a\wedge dA^B$, included in~\sheetextended. 
We therefore see that the Hilbert space of states of the droplet is naturally in representations of the global symmetry group of the full QCD theory. 

With Dirichlet boundary conditions for the gauge field $a$, famously, we find a chiral edge model, $\omega$, moving at a velocity $v$ which does not have to coincide with the speed of light. The chiral edge mode is periodic $\omega\simeq \omega+2\pi$. (In this normalization, the theory of the edge has a factor of $N$ in front.) The physics of this edge mode is very familiar from the fractional quantum Hall effect. Baryon symmetry acts explicitly on $\omega$ by shifting it, $\omega\to \omega+\alpha$. A particularly interesting boundary excitation corresponds to the exponential vertex operator in $\omega$, \eqn\bp{{\cal O}_B=e^{iN\omega}~.}
This is the minimal {\it local} operator that carries baryon charge. In the quantum Hall effect, it is the ``electron operator,'' which creates one unit of un-fractionalized charge on the boundary (namely, it creates an electron on the boundary). The physical meaning of the operator~\bp\ is that it creates a state where $\omega$ winds around the boundary circle. The baryon charge density is ${1\over 2\pi}\del_\phi\omega$ (where the angle on the boundary is denoted $\phi$) and hence the total baryon charge is $B={1\over 2\pi}\int \del_\phi \omega$. 
The baryon charge corresponding to the vertex operator~\bp\ is 1. We can write simple classical field configurations that carry baryon charge, such as
 $\omega=\phi$. Since baryon charge is given by 
$B={1\over 2\pi}\int \del_\phi \omega$ it clearly carries one unit of baryon charge. This classical configuration can be continued to the bulk of the sheet where there is a transparent Dirac string (see~\refs{\MooreYH,\ElitzurNR}).

 The energy of the state corresponding to the operator~\bp\ over the ground state is $\sim N/L$ and the spin of this operator is exactly $N/2$. This value fo the spin precisely matches the expected spin of the minimal baryon made out of one flavor! (Indeed, the spin of the baryon $uuu$, $\Delta^{++}$, is 3/2.) 

Let us clarify how we have computed the spin:  First, we ask how the droplet with the state~\bp\ transforms under rotations that map the disc to itself. This is the same as asking about the $L_0$ eigenvalue of the theory on the boundary (this also happens to be related to the energy gap of the boundary state~\bp\ over the ground state  since the boundary theory is chiral). For this we need to just compute the scaling dimension of the vertex operator~\bp\ in the compact scalar theory on the plane. This comes from the two point function $\langle e^{iN\omega}(x) e^{-iN\omega}(0)\rangle$, which in terms of the complex coordinate $x$ on the plane is given by $x^{-N}$.  This means that the scaling dimension and spin of the vertex operator $e^{iN\omega}(x)$ is $N/2$. Therefore, when we put the theory on the cylinder and if the size of the circle is $L$, the energy stored in the boundary mode would be proportional to $N/L$ and the spin is exactly $N/2$. 

So far we have computed the eigenvalue of the state under rotations that take the disc to itself, $L_0$. Next, we need to embed this configuration in an $SU(2)$ spin representation and clearly the smallest such representation will be of $SU(2)$ spin $N/2$.\foot{Taking into account the ``tumbling motion'' of the disc there will states with higher $SU(2)$ spin. We thank E. Witten for a discussion of this.}

Similarly to the confining string, as soon as there is a boundary, our Hall droplet prefers to shrink due to its tension (and the tension of the vortex loop surrounding the droplet~\vortextension). Since the tension of the boundary as well as the tension of the droplet are both positive and scale linearly in $N$, it is clearly preferable for the droplet to simply disappear. Indeed, without the winding mode~\bp\ the droplet will eventually disappear as it is not protected by any conserved charge. But as soon as the vertex operator $\bp$ is excited, the non-perturbative creation of holes, or the natural tendency of the droplet to shrink due to its tension, cannot eliminate this configuration. The final product is a baryon with size that does not scale with $N$.
More concretely, one can start from a classical configuration which is not an exact solution of the full system of the sheet coupled to the bulk, but the configuration carries baryon number, such as the configuration $\omega=\phi$ that we discussed above. We can be certain that this configuration cannot disappear regardless of the fact that when the droplet becomes small the full set of equations governing the droplet is unknown.

One can understand qualitatively why the excitation corresponding to the baryon vertex operator~\bp\ may stop the droplet from shrinking altogether: In the presence of the baryon vertex operator~\bp\ there is a force that counteracts the tension of the droplet and the tension of the vortex loop. Qualitatively, the energy of a droplet of size $L$ is (omitting unimportant numerical factors) 
\eqn\forces{E(L)=TL^2+T_{v} L+N/L~,} where $T$ is the tension of the droplet (i.e. the tension of the sheet~\tensionsheet) and $T_{v}$ is the tension of the vortex loop, which likewise scales linearly in $N$~\vortextension. These two terms simply prefer to shrink the droplet altogether. But the last term, which is due to the baryon vertex operator's energy above the ground state (it is essentially $L_0/L$) counteracts the tension forces and may lead to a stable droplet with typical size $\Lambda_{QCD}^{-1}$. Of course, by the time the droplet is small one cannot completely trust equation~\forces\ (it is, however, trustworthy for large droplets), but it is encouraging to see that one can understand qualitatively why the droplet may be stable in the full theory in the presence of this vertex operator. Of course, it is guaranteed to be so because of the conservation of baryon number. 

The fact that the baryon settles at a size of order $\Lambda_{QCD}^{-1}$ (and does not scale with $N$) is precisely what we expect from the baryon at large $N$ on phenomenological grounds: while its energy scales linearly in $N$, its size does not. 

We can dress the operator~\bp\ with additional excitations which would not change the total baryon charge. For instance, we can consider the state corresponding to the vertex operator $$\del\omega e^{iN\omega}~.$$ This would be an excitation of the baryon, where the energy larger by $\sim 1/L$, with $L$ the typical size of the baryon, which, as we argued, is $\Lambda_{QCD}$. Therefore, the energy of this excited state differs essentially by $\Lambda_{QCD}$ from the baryon ground state.  Similarly, this configuration has one unit of angular momentum higher than the ground state, so the spin is now $N/2+1$.
It is very tempting to identify this excitation with the following object in the quark model:  we can excite just one of the quarks to the next state, which has a higher spin by one unit. Since each of the quarks carries energy of order $O(N^0)$, we obtain a baryon with spin larger by one unit and the mass increases by a term that scales like $O(N^0)$. This matches qualitatively the physics of the state corresponding to the vertex operator $\del\omega e^{iN\omega}$.

Let us summarize what we have found so far: 

\medskip

\item{1.} QCD in the large $N$ limit admits a meta-stable 2+1 dimensional sheet, which carries a Chern-Simons theory on its world-volume. The  theory on the sheet is nontrivial due to the fact that the $\eta'$ effective theory has a subtle singularity.

\item{2.} The sheet can be coupled to the baryon number gauge field. The excitations of the sheet are de-confined anyons which carry fractional baryon number. The basic anyon can be identified with the microscopic quark (carrying baryon charge $1/N$). The quark therefore is liberated near the sheet and it turns into an anyon. 

\item{3.} Boundary excitations of this sheet are chiral and  certain boundary vertex operators correspond to states with baryon number. 
When these boundary configurations are excited, the sheet cannot completely decay. The excitations that carry baryon number can be simply understood as classical configurations where the compact scalar field winds as we traverse the boundary.

\item{4.} The spin, mass, size, and excitations in this Hall droplet picture agree with the expected properties of QCD baryons.\foot{Interesting previous interpretations of domain wall junctions in terms of heavy particles can be found in \GabadadzePP\ and in the related work~\GabadadzeVW.}

\medskip 

We see that the wave function of the gluons in the baryon we constructed is somewhat exotic: the flux tube breaks near the surface of the baryon and the quarks turn into anyons. In addition, the boundary admits chiral excitations which are gapless for large droplets. While the physical, minimal-energy, baryon is a small droplet and hence the chiral edge modes are not well separated from other excitations, since the droplet is very long lived, we can also imagine (as a thought experiment) pumping energy and spatially stretching the droplet. For instance, we may impose Dirichlet boundary conditions. Then the chiral edge modes are parametrically the lowest lying excitations of the baryon, in addition to the Nambu-Goto excitations of the surface.  In this limit many things can be rigorously computed, in particular, the wave function of the droplet and its quantum numbers. (One may also achieve stretching the droplet by rotating the whole system.)

Let us discuss some aspects of  the physical interpretation of the baryon state we have considered~\bp. The total baryon number at the edge is 1 and therefore we can think about the baryon as a pancake with the quarks all sitting at the boundary. 

But what if we try to push one of the quarks to the bulk of the pancake? Then we should study the Hilbert space of the disc, pierced by the fundamental Wilson line, $P e^{i\int a}$.
The Hilbert space is now a module of the chiral algebra (see~\refs{\MooreYH,\ElitzurNR}). By that we mean that now operators on the boundary do not need to create integer winding for $\omega$. We can have states on the boundary which, for example,  correspond to the operators $$e^{\pm iN\omega-i\omega}~.$$
These have baryon number 1 and -1, respectively. 

The physical meaning of the $B=1$ state above, $e^{i(N-1)\omega}$ is that we have pushed one quark into the pancake and now there are $N-1$ quarks on the boundary. The energy of this state over the ground state of the pierced disc scales like $(N-1)^2/NL$. The ground state of the pierced disc has the state $e^{-i\omega}$ on the boundary and its total baryon number vanishes. 

In pure Chern-Simons theory, the Hilbert space of the pierced disc and the Hilbert space of the empty disc are entirely disconnected objects. Since there are no dynamical anyons but only probe anyons in the theory, the pierced disc Hilbert space can be thought of as the Hilbert space in the presence of a world-line for such a probe particle and therefore it is separated by an infinite energy gap over the Hilbert space of the empty disc. But in the present context, the quarks are not probe particles and it is therefore legitimate to compare the two situations. The ground state of the disc pierced by one quark therefore certainly has an energy higher than the ground state of the empty disc by something that does not scale with $L$ since it comes from (at the very least) the rest energy of the anyon. 

So while by pushing one quark inside the disc we gain a little bit, since the energy gap over the ground state of the pierced disc is now scaling like $(N-1)^2/NL$ (rather than $N/L$), but we also loose because we have to pay some price which is $L$ independent due to the world-line of the bulk anyon. Therefore, in the regime where everything is clearly computable (large $L$), the quarks tend to like to stay all near the boundary. 

This perhaps makes sense in the microscopic picture -- in this way the quarks can overcome the spin-spin repulsion in the most efficient way.  When the droplet's size is of order $\Lambda_{QCD}^{-1}$, we cannot compute all of its properties from first principles. We do not know for sure that the quarks do not drift inside the pancake. But it is tempting to think that what we have found above, namely, that the baryon density is localized near the edge, continues to hold true and the effective field theory description of baryons using the pure Chern-Simons theory is qualitatively correct. (The same point concerns the discussion below of $N_f>1$.)

\newsec{Brief Comments about $N_f>1$}
Let us consider now the $SU(N)$ gauge theory coupled to $N_f$ light Dirac quarks in the fundamental representation. We take $\theta_{QCD}=0$ and we let the quarks have a small positive mass.
At large $N$ the $\eta'$ particle is light and hence the infrared theory is well described by a matrix $U\in U(N_f)$.
The cohomology ring (with integer coefficients) is generated by $Tr((U^{-1}dU)^{2k-1})$ with $1\leq k\leq N_f$. In particular, in four space-dimensions, the baryon current of the Skyrme model corresponds to $\star Tr((U^{-1}dU)^{3})$. This makes sense for $N_f>1$. Since the same current can be written for the $SU(N_f)$ group manifold, one often ignores the $\eta'$ in the construction of Skyrmions. 

Let us however focus on the conserved current $\star Tr(U^{-1}dU)$, which is the direct analog of~\threei. We can think about it as a two-form $U(1)$ symmetry, and it is again unbreakable within the effective theory (and not spontaneously broken in the vacuum). Since the quarks have a small positive mass there is a unique trivial vacuum at $U=\unit$. The sheet which is protected by the symmetry  $\star Tr(U^{-1}dU)$ implements a jump of the matrix $U$ from $\unit$ to $e^{2\pi i } \unit$.
Let us consider some particular interesting sheet which does not support any Nambu-Goldstone bosons on its surface. In order to avoid having Nambu-Goldstone fields trapped to the sheet, along the trajectory of the field $U$, it has to be everywhere a scalar matrix, $U=e^{i\eta'}\unit$, with $\eta'$ changing across the sheet from $0$ to $2\pi$.

The central point regarding this construction is that again while $U$ undergoes this motion, it crosses through singularities which induce a nontrivial TFT on the sheet. It easy to determine the TFT that we should expect on the sheet:  We can again think about the condensate $\langle \Psi \tilde \Psi \rangle\sim N\Lambda^3 U$. If we perform a chiral transformation on $\Psi_i$ and $\tilde\Psi^j$ this corresponds to changing the phase of $U$ along with the mass term. 
By the time the chiral transformation becomes essentially just fermion number, $\Psi_i\to -\Psi_i$, $\tilde \Psi^j\to -\tilde\Psi^j$, the theta angle transforms as $\theta\to \theta+2\pi N_f$. Therefore, the sheet now needs to support the Chern-Simons TFT $SU(N)_{-N_f}$.

For what follows, it will again prove very useful to perform level-rank duality
\eqn\lrdnb{SU(N)_{-N_f}\longleftrightarrow U(N_f)_N~,}
which generalizes~\lrd. If we denote the gauge field of $u(N_f)$ by $a$, then baryon number couples as before through \eqn\baryonco{
\int_{{\cal M}_3}{1\over 2\pi}dA^B\wedge Tr(a)~.} We would like to do more than to just couple the Chern-Simons theory to baryon symmetry. We would like to couple it to the global isospin $SU(N_f)$ symmetry, since that would allow us to read out the quantum numbers of our edge modes under isospin! This would be a new nontrivial consistency check of the picture as we could then compare the quantum numbers of the edge modes to those of the microscopic baryons. 

Note that~\baryonco\ implies that some Wilson lines carry fractional baryon charge. Hence, the $U(1)_B$ symmetry is fractionalized on the sheet.  On the other hand, the $SU(N_f)$ quantum numbers are, of course, not fractionalized. 
This is what we expect in terms of the quark model: the quarks carry fractional baryon number ($1/N$ for the basic quark) but they are in the fundamental representation of isospin. 

A quick way to understand the representation theory of the Hall droplet is as follows. At first sight, it is somewhat confusing that the {\it gauge symmetry } of our TFT, $U(N_f)_N$, is $U(N_f)$ but also the {\it global symmetry} of the problem (including baryon number and isospin) is $U(N_f)$. 
This is actually a very happy coincidence since on the edge of the sheet the gauge symmetry becomes a global symmetry (this happens due to the Dirichlet boundary conditions, as in the previous section). Consequently, the Hilbert space of the droplet naturally furnishes a representation of baryon number symmetry and isospin, exactly as required. The fact that there is such a simple and natural boundary condition for the TFT that leads to states in representations of the symmetry group of QCD should be viewed as nontrivial consistency check of the whole picture.

Next, it is evident that if we consider a transparent Wilson line then it may end on the boundary and it leads to a genuine local operator on the boundary. In our discussion of the case $N_f=1$, this was the Wilson line $P e^{i N \int a}$, which led to the vertex operator~\bp\ on the boundary. 
We can follow a similar strategy here and ask which Wilson lines in the TFT $U(N_f)_N$ are transparent. What we find is that the Wilson line in the representation $Sym^N(\square)$ of the gauge symmetry $U(N_f)$ is transparent.\foot{This fact was also recently employed in the literature on 2+1 dimensional dualities~\refs{\AharonyMJS,\BeniniAED}.} In other words, the Wilson line
$$Tr_{Sym^N(\square)} P e^{i \int a }~,$$ 
with $a$ a $u(N_f)$ gauge field is a transparent line in the theory.
We can now read out the quantum numbers of the corresponding boundary state. Since the gauge symmetry becomes a global symmetry on the boundary,
we immediately see that the baryon number is 1, and the representation under isospin symmetry is $Sym^N(\square)$. 
We can also ask about the spin of boundary state. This requires a little bit more work, namely, we need to calculate the $L_0$ eigenvalue of this boundary state. This comes out to be precisely $N/2$ (to obtain this result one uses standard results about the chiral algebra corresponding to this $U(N_f)_N$ Chern-Simons theory). 

We therefore see that we have described a state in the theory, which carries spin $N/2$, it is in the $Sym^N(\square)$ representation of isospin, and has baryon number 1. This state has the baryon density localized near the boundary.

Let us contrast this with the baryons in the microscopic, 3+1 dimensional, theory. We studied them in~\symmrep. 
We see that what we find with the droplet model precisely matches the baryon in the theory with $N_f>1$ that is made out of one flavor (and its friends under isospin -- in QCD, where $N=3$, we are describing the famous decuplet, whose members have spin $3/2$). In addition, the mass and size of the baryon also agree with the expected scaling. It is hard to imagine that this detailed agreement is a coincidence. 

Note that we have again only calculated the spin in the sense of the eigenvalue of the rotations in the plane of the sheet and this needs to be embedded in an $SU(2)$ representation. But clearly the smallest such $SU(2)$ representation has spin $N/2$. 

So far we have only discussed the multiplet of baryons of spin $N/2$. In the future, the other baryons need to be described within this framework as well. This would allow for a detailed analysis of the spectroscopy of baryons and comparison with~\AdkinsYA.

A final question to discuss concerns with the connection of the Hall droplet picture with the Skyrmion model, which employs the conserved current $\star Tr((U^{-1}dU)^{3})$ and essentially ignores the $\eta'$ dynamics. The above computations lead to the tempting picture that the $\eta'$ dynamics, the associated Hall droplet and its chiral edge states describe well the high-spin baryons and for the low-spin baryons the ordinary Skyrmion picture is more appropriate. Note that since there is a unique baryon symmetry, the $A^B$ gauge field in~\baryonco\ is the same one that couples in the bulk (i.e. away from the sheet) to $\star Tr((U^{-1}dU)^{3})$. There should be transitions between the ordinary Skyrmions and our Hall droplets.  It would be nice to understand how this works in detail. 

In the real world, of course, $N=3$ and therefore the continuation to large $N$ of various baryons is ambiguous. For instance, consider the $\Delta^{++}$ or $\Omega^{-}$ baryons. They have spin 3/2. We can think at large $N$ either about baryons that have spin $3/2$ (and are made out of more than one quark flavor) or about baryons that are made of just one flavor and have spin $N/2$. These are both valid large $N$ extrapolations of such baryons. 
In the former case the ordinary Skyrmion picture makes sense and the $\eta'$ can be ignored. But in the latter case we presumably have the Hall droplet model, which as we have seen above, describes the quantum numbers and some of the properties of the $\Delta^{++}$ and $\Omega^{-}$ baryons correctly. 
These two points of view may illuminate different aspects of such baryons. For example, a famous result from the Skyrmion model is that the mass splitting between the spin $3/2$ and spin $1/2$ baryons scales like $O(1/N)$~\AdkinsYA. But we can also extrapolate the decuplet and octet to large $N$ as spin $N/2$ and spin $N/2-1$ baryons, respectively. Then our model suggests that the mass difference scales like $O(1)$. This may better fit the phenomenology, as the mass difference between the decuplet and octet is famously somewhat off in the Skyrmion model.

\bigskip
\noindent{\bf Acknowledgments}

We would like to thank A.~Abanov, O.~Aharony,  R.~Argurio, A.~Armoni, A.~Cherman, C.~Choi, C.~Cordova, J.~Gomis, B.~Gudnason, P.~Niro, E.~Poppitz, N.~Seiberg, M.~Shifman, M.~Strassler, E.~Witten.
We would also like to especially thank M.~Karliner. Z.K. is supported in part by the Simons Foundation grant 488657 (Simons Collaboration on the Non-Perturbative Bootstrap). Any opinions, findings, and conclusions or recommendations expressed in this material are those of the authors and do not necessarily reflect the views of the funding agencies.

\listrefs

\end